\documentclass[11pt]{article}

\usepackage{amsmath}
\usepackage{amssymb}
\usepackage{mathptmx}
\usepackage{amsfonts}
\usepackage[mathscr]{eucal}
\usepackage[dvips]{graphicx}

\newcommand{\mbf}[1]{\ensuremath{\mathbf{#1}}}
\newcommand{\ms}[1]{\ensuremath{\mathscr{#1}}}

\newcommand{\sqzb}[1]{\ensuremath{d\Gamma_{\textrm{b}}({#1}) }}
\newcommand{\sqzf}[1]{\ensuremath{d\Gamma_{\textrm{f}}({#1}) }}
\newcommand{\domain}[1]{\ensuremath{\mathscr{D} ({#1})}}

\newcommand{\tens}{\otimes}
\newcommand{\ntens}{\otimes^{n}}
\newcommand{\nstens}{\otimes^{n}_{s}}
\newcommand{\natens}{\otimes^{n}_{a}}
\newcommand{\half}{\frac{1}{2}}
\newcommand{\fint}{\psi^{\ast}(\mathbf{x})\alpha^{j}\psi
(\mathbf{x})}

\newcommand{\piE}{( 2 \pi )^{3}E_{M} }
\newcommand{\eltwo}{L^{2}(\mathbf{R}^{3})}
\newcommand{\Rthree}{\mathbf{R}^{3} }
\newcommand{\dx}{d \mathbf{x} }
\newcommand{\dy}{d \mathbf{y} }
\newcommand{\dk}{d \mathbf{k} }

\newcommand{\restr}{\upharpoonright}
\newcommand{\core}{\ms{F}^{\textrm{fin}}_{\textrm{el}} (\ms{D}(E_{M}))
 \hat{\otimes} \ms{F}^{\textrm{fin}}_{\textrm{ph}}  (\ms{D}(\omega))}
\newcommand{\coremV}{\ms{F}^{\textrm{fin}}_{\textrm{el}} (\ms{D}(E_{M}))   
 \hat{\otimes} \ms{F}_{\textrm{ph}}^{\textrm{fin}}  (\ms{D}(\omega_{m,V}))}

\newcommand{\I}{\textrm{I}}
\newcommand{\II}{\textrm{II}}
\newcommand{\ph}{\textrm{ph}}
\newcommand{\el}{ \textrm{el} }
\newcommand{\bos}{\textrm{b}}
\newcommand{\fer}{\textrm{f}}
\newcommand{\fin}{\textrm{fin}}
\newcommand{\QED}{\textrm{QED}}

\newtheorem{theorem}{Theorem}[section]
\newtheorem{proposition}[theorem]{Proposition}
\newtheorem{lemma}[theorem]{Lemma}

\newtheorem{remark}{Remark}[section]

\begin{document}

\title{On the Spectral Analysis of Quantum \\ 
 Electrodynamics with Spatial Cutoffs. I}
\author{Toshimitsu TAKAESU}
\date{ }
\maketitle
\begin{center}
\textit{Faculty of Mathematics, Kyushu University, \\
 Fukuoka, 812-8581, Japan }
\end{center}

\begin{quote}
\textbf{Abstract.}
In this paper, we consider  the spectrum of a model
in quantum electrodynamics with a spatial cutoff.
   It is proven that (1) the Hamiltonian is self-adjoint;
   (2) under the infrared regularity condition, the Hamiltonian has a unique ground state for sufficiently small values of coupling constants.
 The spectral scattering theory is studied as well
 and
   it is shown that  asymptotic fields exist  and
 the spectral gap is closed.
 \end{quote}

\section{Introduction}
The present paper investigates
the existence and uniqueness of the ground state of a model in quantum electrodynamics (QED) in the Coulomb gauge. QED has, of course, been studied so far from the physical point of view. Nevertheless, it is interesting to investigate it purely from the mathematical standpoint. Indeed it is 
 not so well understood in mathematical rigor.

The first task is to realize the Hamiltonian
of QED as a self-adjoint operator on an appropriate Hilbert space, which
means that the Hamiltonian generates a unique unitary time evolution. What we need to do mathematically
 is to
specify conditions under which the Hamiltonian is a self-adjoint operator.
The second task is to study the spectral properties of the Hamiltonian defined as a self-adjoint operator.
The eigenvector associated with the bottom of the spectrum of
a self-adjoint operator is called {\it ground state} if it exists.
We are concerned with the ground states of our self-adjoint operator
associated with a model in QED.
Note that it is not  trivial to show the existence and uniqueness of the ground state, since the ground state for zero coupling is
embedded in the continuous spectrum and then
the regular perturbation theory \cite{Kato} does not work directly even for nonzero but sufficiently weak couplings.   Nevertheless, we can give a sufficient condition such that
      the unique ground state exists.

Before starting a rigorous discussion, we roughly review
our model for readers' convenience.
  Informally, the standard Hamiltonian of QED in
 the Coulomb gauge {\it without} external potentials is given by
\begin{equation}
\label{defh}
H=H_{\el}+H_{\ph}
+\alpha  \int_{\Rthree}
\psi^{\dagger}(\mbf{x} ) \alpha^{j} \psi (\mbf{x} ) A_{j}(\mbf{x} ) d\mbf{x} +
\alpha^2   \int_{\Rthree \times \Rthree}
\frac{
\psi^{\dagger}(\mbf{x} ) \psi (\mbf{x} )
   \psi^{\dagger}(\mbf{y} ) \psi (\mbf{y} )}
   {|\mbf{x} -\mathbf{y}|}
      d \mbf{x} \, d\mathbf{y} ,
\end{equation}
where $\alpha\in R$ denotes the  coupling constant,
$\psi$  the Dirac field, $A_j$ the quantized radiation field, $\alpha^j$, $j=1,2,3$,  $4\times4$ Dirac matrices satisfying canonical anticommutation relations,
and $H_{\el}$ and $H_{\ph}$
the kinetic term of the Dirac field and the photon field, respectively. Here, ultraviolet cutoffs $\chi_{\ph}$ and $\chi_{\el}$ are imposed on
$A^j$ and $\psi$, respectively.
All the definitions are given rigorously in
Section 2.
The first two terms on the right-hand side of (\ref{defh}), \begin{equation}
H_{\el}+H_{\ph},
\end{equation}
  describe the zero coupling Hamiltonian which is
  a well defined nonnegative self-adjoint operator on the tensor product of a Fermion Fock space and a Boson Fock space, $\ms{F}_{\QED}$,  and
the bare vacuum of $ \ms{F}_{\QED}$
is its unique ground state.
While the third term describes the minimal coupling between the Dirac field and the quantized radiation field, the last term is derived from
the Coulomb gauge condition.

Although (\ref{defh}) is a standard Hamiltonian
in QED in the Coulomb gauge,
it is not clear if $H$ with $\alpha\not=0$
is well defined and can be realized
as a  self-adjoint operator bounded from below.
One of the simplest ways to realize $H$ as a well defined self-adjoint operator is
to
introduce a real spatial cutoff function $\chi$,
and thus, we define the spatial cutoff Hamiltonian
by replacing the Dirac field $\psi$ with $\chi\psi$.
Namely our Hamiltonian turns out to be of the form
\begin{align}
H_{\chi}=
H_{\el}+H_{\ph}
&+\alpha  \int_{\Rthree}
\chi(\mbf{x} )
\psi^{\dagger}(\mbf{x} ) \alpha^{j} \psi (\mbf{x} ) A_{j}(\mbf{x} ) d\mbf{x} \nonumber \\
&\label{defh2}+
\alpha^2   \int_{\Rthree \times \Rthree}
\chi(\mbf{x}) \chi(\mbf{y})
\frac{
\psi^{\dagger}(\mbf{x} ) \psi (\mbf{x} )
   \psi^{\dagger}(\mbf{y} ) \psi (\mbf{y} )}
   {|\mbf{x} -\mathbf{y}|}
      d \mbf{x} \, d\mathbf{y} .
\end{align}
Suppose that $\chi$ satisfies that
\begin{eqnarray}
\label{self}
\int_{\Rthree}  | \chi(\mbf{x})  | dx<\infty \qquad 
 \mbox{ and } \qquad 
\int_{\Rthree \times \Rthree}  
\frac{| \chi(\mbf{x}) \chi(\mbf{y}) |}{|\mbf{x}- \mbf{y}|}  \dx \dy <  \infty.
\end{eqnarray}
By virtue of (\ref{self}) it can be seen that the interaction term
in (\ref{defh2})  is  a well defined symmetric operator and
 moreover $H_{\chi}$ is a self-adjoint operator
 bounded from below for all $\alpha\in R$.
$H_\chi$ is the main object in this paper.

$\quad $ \\
Next, we study the spectral properties of $H_{\chi}$. \\ 
(\textit{Translation invariance})
Since $H$ has no external potential,
it is translation-invariant, i.e., $H$ is invariant
under transformation
$$\psi(\mbf{x})\rightarrow\psi(\mbf{x}+\mbf{a}), \quad
A_j(\mbf{x}) \rightarrow A_j(\mbf{x}+\mbf{a})$$ for arbitrary $ \mbf{a} \in \Rthree$.
  It is crucial, however, that the spatial cutoff breaks 
  this translation invariance. 
      Thus, the cutoff could be regarded as an external potential.  \\
 \textit{(Ground state)}   
      In addition to (\ref{self}) supposing integrability:
\begin{eqnarray}
\label{existence}
 \int_{\Rthree} |\mbf{x} | |\chi(\mbf{x}) | \dx <\infty,
\end{eqnarray}
we can show that
   $H_{\chi}$ has a unique ground state for a sufficiently small coupling constant under
   the infrared regularity condition :
\begin{equation}
\int_{\Rthree}
\frac{ |\chi_{\ph}(\mbf{k})|^2}{\omega(\mbf{k})^3} \dk <\infty,
\end{equation}
where $\omega(k)=|k|$ denotes the dispersion relation of photons. \\
{(\textit{Total charge})}
Let
\begin{eqnarray}
Q=N_+   -N_-
\end{eqnarray}
 be the total charge of the Dirac field, where $N_+$ (resp. $N_-$)
denotes the number operator for electrons  (resp. positron).
Since $\psi^\dagger (\mbf{x}) \psi(\mbf{x})$ leaves the total charge invariant for each $x  \in \mbf{R}^3$;
 $ e^{ it Q}  H_{\chi}  e^{-itQ} =H_{\chi} $, 
  $H_\chi$ leaves the total charge invariant.
Then, $ \ms{F}_{\QED}$ is decomposed
with respect to the spectrum of the total charge as
\begin{eqnarray}
\ms{F}_{\QED}=\bigoplus_{z \in \mathbb Z} \ms{F}_z.
\end{eqnarray}
It can be shown that
the unique ground state of $H_\chi$ belongs to $\ms{F}_0$, i.e., the total charge of the ground state is zero.

Finally, we also establish that $H_\chi$ has no spectral gap,
i.e., the gap between the bottom of the spectrum and that of the continuum is closed.
This is established by constructing asymptotic fields.

The main roles of the spatial cutoff are:
\begin{itemize}
\item[(1)] it ensures well defined self-adjointness of the Hamiltonian;
\item[(2)] it leaves  the total charge invariant;
\item[(3)] it breaks translation invariance and serves as an external potential.
\end{itemize}
The spectral analysis of this kind of system
has been developed  in the last decade, and many results have been obtained.  
In particular, this paper is inspired by
       Arai-Hirokawa \cite{AH97}, where
       generalized spin-boson (GSB) models  are studied.
 Dimassi-Guillot \cite{DiGu03} and Barbaroux-Dimassi-Guillot \cite{BDG04} also study QED, in which
  the Hamiltonian has an external potential.  In \cite{DiGu03},
  \cite{BDG04}
    the self-adjointness of the Hamiltonian,
    and the existence and uniqueness of the ground state are established under certain conditions.
Furthermore, Bach-Fr\"{o}hlich-Sigal \cite{BFS99},
      Sphon \cite{Sp98}, G\'{e}rard \cite{Ge00}, and Grisemer-Lieb-Loss \cite{GLL01} and references therein discuss related models.

Finally, we provide several technical comments comparing  with the GSB models studied in \cite{AH97}.
\begin{description}
\item[(Existence of the ground state)]
In order to
prove the existence of the ground state, we use the momentum lattice approximation.
References \cite{BFS99}, \cite{Hi05} prove the existence of the ground state by combining the spatial localization
of nonrelativistic electrons and the momentum lattice approximation.
  In our case,
the spatial localization is converted into  the assumption
$\int_{\Rthree} |\mbf{x}| \chi(\mbf{x}) \dx < \infty$.
Note that the interaction terms in
 GSB models are of the simple form $\sum_{j=1}^N A_j\otimes B_j$.
Thus, the
localization argument is not needed in GSB models.
\item[(Uniqueness of the ground state)]
The physically realistic
  dispersion relation is $\omega(\mbf{k})=|\mbf{k}|$. Namely,
photons are massless.
Thus, in showing the uniqueness of the ground state,
the min-max principle applied in \cite{AH97} for massive boson is not applicable. Instead, we apply the strategy given in \cite{Hi06}.

\item[(Non-compact resolvent)]
 In \cite{AH97}, the fermion term of the zero coupling Hamiltonian
is assumed to have a compact resolvent.
In our case, however, $H_{\el}$ has no compact resolvent since $\sigma(H_{\el})=\{0\}\cup [m,\infty)$.
So, we apply the strategy given in \cite{BFS99},
 \cite{Hi01}, \cite{Hi05} for nonrelativistic QED.

\item[(Asymptotic fields)] In \cite{H-K68}, the asymptotic field is constructed 
for massive cases.
    However, since we have to cover massless cases, we construct it through the stationary phase method discussed in \cite{FGS01},  \cite{Hi01}.
\end{description}

This paper is organized as follows: 
 Section 1 is devoted to defining
 the Hamiltonian with spatial cutoffs,  which is a slight generalization of $H_\chi$ mentioned above,  and
 stating the main results.
      In Section 2, we prove that for sufficiently small values of the coupling constant, a unique ground state of the Hamiltonian exists under
 the  infrared regularity condition.
 In section 3,  the spectral scattering theory is considered and
 it is shown that the spectral gap is closed.
 Section 4 proves that the total charge of the ground state is zero.

\subsection{Boson Fock Spaces and Fermion Fock spaces}
  Let $\ms{X}$ and $\ms{Y}$ be  Hilbert spaces over $\mbf{C}$.
We denote by $\nstens \ms{X}$ the n-fold symmetric tensor product of $\ms{X}$ and by  $\natens \ms{Y}$ the n-fold anti-symmetric tensor product of $\ms{Y}$.
The boson Fock space over $\ms{X}$ is defined by
\[
\ms{F}_{\bos}(\ms{X} ) := \oplus_{n=0}^{\infty}
 (  \nstens \ms{X} ) 
  := \left\{ \Psi = \{ \Psi^{(n)} \}_{n=0}^{\infty} 
 \left|  \frac{}{} \right. \Psi^{(n)} \in \nstens \ms{X} , \, \sum_{n=0}^{\infty} \|  \Psi^{(n)}  \|_{ \otimes^{n} \ms{X} }^{2} < \infty \right\} ,
\]
and the fermion Fock space over $\ms{Y}$ by  
 \[
\ms{F}_{\fer}(\ms{Y} ) := \oplus_{n=0}^{\infty}
 (  \natens \ms{Y} ) 
  := \left\{ \Psi = \{ \Psi^{(n)} \}_{n=0}^{\infty} 
 \left|  \frac{}{} \right. \Psi^{(n)} \in \natens \ms{Y} , \, \sum_{n=0}^{\infty} \|  \Psi^{(n)}  \|_{ \otimes^{n} \ms{Y} }^{2} < \infty \right\} .
\]
$ \ms{F}_{\bos}(\ms{X} )$ is the Hilbert space with 
the inner product $ 
(\Phi, \Psi )= \sum_{n=0}^{\infty}
( \Phi^{(n)} , \; \Psi^{(n)} )_{ \otimes^{n} \ms{X} }  $, 
and also $  \ms{F}_{\fer}(\ms{Y} )$ is the Hilbert space with
 the same inner product.
In this paper, the inner product $( g , f  )_{\ms{K}}$  of 
 Hilbert space $\ms{K} $ is linear in $f$ and anti-linear in $g$.
Let $S_{n} : \ntens \ms{X} \to \nstens \ms{X}  $ and
$A_{n} : \ntens \ms{Y} \to \natens \ms{Y} \; $ be  orthogonal projections. 
For $\xi \in \ms{X}$, the creation operator $\; A^{\ast}(\xi )\;$ 
 on  $\ms{F}_{\bos} (\ms{X}) $ is defined by
\[
( A^{\ast}(\xi )\Psi )^{(n)} = 
\sqrt{n+1} S_{n+1} ( \xi \tens \Psi^{(n)} ), \quad n \geq 1 ,
\] 
and $( A^{\ast}(\xi )\Psi )^{(0)} = 0  $,  
 while the creation operator $\; B^{\ast}(\eta )\; $ on $\ms{F}_{\fer} (\ms{Y}) $ is  defined by
\[
( B^{\ast}(\eta )\Psi )^{(n)} = 
\sqrt{n+1} A_{n+1} ( \eta \tens \Psi^{(n)} ), \quad n \geq 1 ,
\] 
and $( B^{\ast}(\eta )\Psi )^{(0)} = 0  $.
 The annihilation operators $\; A(\xi) \; $ and
 $ \; B(\eta) \; $ are defined by the adjoint operators of 
 $\; A^{\ast}(\xi )\; $ and $\; B^{\ast}(\eta )\;$, respectively.
Let 
$\Omega_{\bos} = \{ 1,0, \cdots \} \in  \ms{F}_{\bos}( \ms{X} ) $ 
and 
$\Omega_{\fer} = \{ 1,0, \cdots \} \in  \ms{F}_{\fer}( \ms{Y} )$.
We denote the boson-finite particle subspace over $\ms{M} \subset \ms{X}  $  by 
\[
\ms{F}_{\bos}^{\fin}(\ms{M} )= \ms{L} \{  A^{\ast} (\xi_{1}) \cdots  A^{\ast} (\xi_{n})
\Omega_{\bos}, \, \, \Omega_{\bos} \;| \; \xi_{j} \, \in 
\ms{M} , j= 1, \cdots ,n, \, \,  n  \in  \mbf{N} \} ,
\]
and the fermion-finite particle subspace over  $\ms{N} \subset \ms{Y}  $ by 
\[
\ms{F}_{\fer}^{\fin}(\ms{N} )= \ms{L} \{  B^{\ast} (\eta_{1}) \cdots  B^{\ast} (\eta_{n})
\Omega_{\fer}, \, \, \Omega_{\fer} \;| \; \eta_{j} \, \in 
\ms{N} , j= 1, \cdots ,n, \, \,  n  \in  \mbf{N} \} .
\]
For simplicity, we call $ \ms{F}_{\bos}^{\fin}( \ms{X})$ the 
finite particle subspace.
The domain of $A^{\ast} (\xi )$ is given by
\[
\domain{ A^{\ast} (\xi ) }  
=  \{ \Psi = \{ \Psi^{(n)} \}_{n=0}^{\infty} 
 \in \ms{F}_{\bos}
 (\ms{X})    \left| \frac{}{} \right. 
\sum_{n=0}^{\infty} \| ( A^{\ast}(\xi ) \Psi )^{(n)}  \|_{ \otimes^{n} \ms{X} }^{2} < \infty \} .
\]
It is seen that the domains of $A^{\ast}(\xi )$ and 
$A(\xi ' )$ include the finite particle subspace, and leave it 
 invariant. They satisfy the canonical commutation relations on the finite particle space:
\[  [ \, A(\xi ), \, A^{\ast} (\xi ')  ] 
= ( \xi , \xi ')_{\ms{X}} ,  \qquad 
    [ \, A(\xi ), \, A(\xi ')  ] = [ A^{\ast}(\xi ), \, 
 A^{\ast}(\xi ') ] =0 ,
\]
where $[A,B] =AB-BA$.
On the other hand, $B^{\ast}(\eta )$ and 
$B(\eta ' )$ are bounded on $\ms{F}_{\fer}(\ms{Y})$, and satisfy the canonical anti-commutation relations on $\ms{F}_{\fer}(\ms{Y})$:
\[  \{ \, B(\eta ), \, B^{\ast} (\eta ')  \} 
= ( \eta  , \eta ' )_{\ms{Y}} ,  \qquad 
    \{ \, B(\eta  ), \, B(\eta ' )  \} 
 = \{ B^{\ast}(\eta ), \,  B^{\ast}(\eta ' ) \} =0,
\]
where $ \{ A , B \}  = AB +BA  $.
$\quad$ \\ 
Let $X$ be an operator on $\ms{X}$
The  second quantization operator $ \sqzb{X}  \; $ on $\; \ms{F}_{\bos}(\ms{X}) $ is defined by
\[
 \sqzb{X}  = \oplus_{n=0}^{\infty} \left( 
\sum_{j=1}^{n} ( I \tens \cdots I \tens \underbrace{X}_{jth} \tens I   \cdots  \tens I )   \right) .
\]
In the same way as $\sqzb{X}$, we define  $\sqzf{Y} $ 
on $\ms{Y}$ for an operator $Y$ .

\subsection{Radiation Fields}
In this paper, we consider the photon field quantized in the Coulomb gauge.
$\quad $ \\ 
Let 
\begin{equation}
 \ms{F}_{\ph} = \ms{F}_{\bos} 
(L^{2}(\mbf{R}^{3}; \mbf{C}^{2} )),
\end{equation} 
 and  $ \mbf{e}_{r}(\mbf{k} )
= (e_{r}^{j}(\mbf{k} ))_{j=1}^{3} $, $r=1,2$, 
 be the polarization vector satisfying
\begin{equation}
 \mbf{e}_{r}(\mbf{k} ) \cdot \mbf{e}_{r'}(\mbf{k} )=
 \delta_{r,r'} , \quad  
\mbf{k} \cdot  \mbf{e}_{r}(\mbf{k} ) , \quad 
\text{a.e.} \;  \mbf{k} \in \Rthree.
\end{equation}
For $\;f \in \eltwo  $,  let
$ a^{\ast}_{1}(f) = A^{\ast}( (f,0 )) \; $ and    
 $\; a^{\ast}_{2}(f) = A^{\ast}((0,f)) $. 
Then, $a_{r}(f)$ and  $a_{r}^{\ast}(g)$ satisfy the canonical commutation relations:  
\begin{align*}
   & [ \, a_{r}(f ), \, a^{\ast}_{r'} (g )  ] = \delta_{r,r'} (f, g  ), \\
   & [ \, a_{r}(f ), \, a_{r'}(g )  ] = [ a_{r}^{\ast}(f ), \,  a_{r'}^{\ast}(g ) ] =0 ,
\end{align*}
on the finite particle subspace.
The energy of a   photon with momentum 
 $\mbf{k} \in \Rthree $ is given by
\begin{equation}
\omega (\mbf{k} )= | \mbf{k} | .
\end{equation}
The free Hamiltonian of the  photon field  is defined by
\begin{equation}
H_{\ph} = d\Gamma_{\bos} (\omega ) .
\end{equation}
Let
\begin{equation}
f_{r, \mbf{x}}^{j}(\mbf{k}) = \frac{\chi_{\ph} (\mbf{k})e^{j}_{r, \mbf{x}}(\mbf{k}) }{\sqrt{2(2\pi )^{3} \omega (\mbf{k})}},
\end{equation}
where 
$ \; e^{j}_{r, \mbf{x}} (\mbf{k} ) = e_{r}^{j} (\mbf{k} ) e^{-i \mbf{k} \cdot \mbf{x}}$. 
The quantized radiation field  is defined  by
\begin{equation}
A_{j}(\mbf{x}) = \sum_{r=1,2}( a_{r}( f_{r,\mbf{x}}^{j}) +
 a^{\ast}_{r}( f_{r, \mbf{x}}^{j})) .
\end{equation}
We assume the following conditions. 
\begin{quote}
\textbf{(A.1) (Ultraviolet cutoff for the photon 
 field)     } 
\[
\int_{\Rthree} \left| \frac{\chi_{\ph} (\mbf{k}) }{\sqrt{\omega (\mbf{k}) }} \right|^{2} d \mbf{k}  \; <  \; \infty 
 \quad  \text{and}  \quad
\int_{\Rthree} \left| \frac{\chi_{\ph} (\mbf{k}) }{
\omega (\mbf{k}) } \right|^{2} d \mbf{k}  \; <  \; \infty .
\]
\end{quote}
Let $f \; \in  \ms{D} (\omega^{-1/2} )$.
 It is seen that $ a_{r}(f)  $ and $ a_{r}^{\ast}(f) $ are relatively bounded with respect to $H_{\ph}^{1/2}$, i.e., for $\Psi \in \ms{D} (H_{\ph}^{1/2}) $,
\begin{align}
& \| a_{r}(f) \Psi  \| \leq \| \frac{f}{\sqrt{\omega}} \| \, 
\|   H^{1/2}_{\ph} \Psi \| , \label{boundar} \\
& \| a_{r}^{\ast}(f) \Psi  \| \leq \| \frac{f}{\sqrt{\omega}} \| \, 
\|   H_{\ph}^{1/2} \Psi \| + \| f \| \| \Psi \| . 
\label{boundadr}
\end{align}
Then  $ A_{j}(\mbf{x} ) $ is also relatively bounded with respect to $ H_{\ph}^{1/2}  $ :  
 \begin{equation}
\| A_{j}(\mbf{x} ) \Psi \| \leq
\sum_{r=1,2} ( 2 M_{2 ,j,  \, r}^{\ph} \| H_{\ph}^{1/2} \Psi \| + 
M_{1, j, \, r}^{\ph} \| \Psi \|  ) , \label{boundAj}
\end{equation}
where $ \; M_{k, j, \, r}^{\ph} =\left\| \frac{\chi_{\ph}e^{j}_{r}}{\sqrt{2(2\pi )^{3} 
 \omega^{k}}} \right\|  \;  $
     for  $ \; k=1,2, \; r=1,2 $ and $  \; j=1,2,3$. \\

\subsection{Dirac Fields}
Next, we define the Dirac field. Let
\[
\ms{F}_{\el} = \ms{F}_{\fer} (L^{2}(\mbf{R}^{3}; \mbf{C}^{4} )).
\]
The energy of an electron with momentum $\mbf{p}$ 
 is given by
\[
E_{M} (\mbf{p} )= \sqrt{ M^{2} + \mbf{p}^{2} } , \quad \quad M > 0 ,
\]
where $M$ denotes the mass of an electron and we fix it. 
The free Hamiltonian of the Dirac field  is defined by
\begin{equation}
H_{\el} = d\Gamma_{\fer} (E_{M} ) ,
\end{equation} 
Let  
\[
h_{D} (\mbf{p}) = \mbf{\alpha} \mbf{\cdot}
  \mbf{p}   + \beta M  , 
\qquad  s(\mbf{p}) =  \mbf{s} \mbf{\cdot} \mbf{p} ,
\]
where  $\alpha^{j} $, $j=1,2,3$ and  $\beta$ be the $4\times 4$ matrix satisfying the canonical anti-commutation relation: 
\begin{equation}
 \{ \alpha^{j} , \alpha^{l} \} = 2 \delta_{j,l} , \qquad 
\{ \alpha_{j} , \beta \} = 0 ,   \qquad
\beta^{2}=I  , 
\end{equation}
 and   $ \mbf{s} =(s_{j})_{j=1}^{3} $ denotes the angular momentum of spin. Throughout this paper, we fix them. 
   The spinors  $ u_{s}(\mbf{p} ) = 
(u_{s}^{l} (\mbf{p}) )_{l=1}^{4} \; $ describe the positive energy part with spin $s$ and  
 $v_{s}(\mbf{p} ) = (v_{s}^{l} (\mbf{p}) )_{l=1}^{4}  \; $
the negative energy part with spin $s$,  
$s= \pm 1/2 $ :    
\begin{align*}
&h_{D} (\mbf{p}) u_{s} (\mbf{p}) = E_{M} (\mbf{p}) u_{s} (\mbf{p}), 
\qquad    s(\mbf{p} ) u_{s} (\mbf{p}) = s | \mbf{p} | u_{s} (\mbf{p}),   \\ 
&h_{D} (\mbf{p}) v_{s} (\mbf{p}) = 
-E_{M} (\mbf{p}) v_{s} (\mbf{p}), 
\qquad s(\mbf{p} ) v_{s} (\mbf{p}) = s | \mbf{p} | v_{s} (\mbf{p}) .
\end{align*}
These form an orthogonal base of $\mbf{C}^{4}$ 
\[
u_{s} (\mbf{p} )^{\ast} u_{\tau} (\mbf{p}' )
=  v_{s} (\mbf{p} )^{\ast} v_{\tau} (\mbf{p} ' ) 
=\delta_{s,\tau}  \sqrt{E_{M} (\mbf{p})} 
  \sqrt{E_{M} (\mbf{p} ')}  ,\qquad
u_{s} (\mbf{p} )^{\ast} v_{\tau} (\mbf{p}' )
=  v_{s} (\mbf{p} )^{\ast} u_{\tau} (\mbf{p} ') = 0 . 
\]
Moreover, the completeness condition is satisfied 
\[
\sum_{s= \pm 1/2} 
\left( u_{s}^{l} (\mbf{p}) u_{s}^{l'} (\mbf{p})^{\ast}
 +  v_{s}^{l} (\mbf{p}) v_{s}^{l'} (\mbf{p})^{\ast} \right)
= \delta_{l,l'}.
 \]
 Let us set the creation operators by
\begin{align*}
&b_{1/2}^{\ast}(f) = B^{\ast}({}^{t} (f,0,0,0) ), \quad \quad  
b_{-1/2}^{\ast}(f) = B^{\ast}({}^{t}(0,f,0,0) ), \\
&d_{1/2}^{\ast}(f) = B^{\ast}({}^{t}(0,0,f,0) ), \quad \quad  
d_{-1/2}^{\ast}(f) = B^{\ast}({}^{t}(0,0,0,f) ) , 
\end{align*}
for $f \in \eltwo$. 
Then, the creation and annihilation operators satisfy the CAR: 
\begin{align*}
&\{ b_{s} (f) , b_{\tau}^{\ast}(g) \} = 
\{ d_{s} (f) , d_{\tau}^{\ast}(g) \} =
\delta_{s, \tau} (f, g)   ,  \\
&\{ b_{s} (f) , b_{\tau} (g) \} = 
\{ d_{s} (f) , d_{\tau} (g) \} = 0   , \\
& \{ b_{s} (f) , d_{\tau} (g) \} = 
\{ b_{s} (f) , d_{\tau}^{\ast}(g) \} = 0  .
\end{align*}
Let
\begin{equation}
g_{s, \mbf{x}}^{l} (\mbf{p}) = 
\frac{\chi_{\el}(\mbf{p} ) u_{s,\mbf{x}}^{l} (\mbf{p}) }{\sqrt{\piE (\mbf{p})}}     , \qquad  
h_{s, \mbf{x}}^{l} (\mbf{p})= 
\frac{\chi_{\el} (\mbf{p}) \tilde{v}_{s, \mbf{x}}^{l} (\mbf{p})}
{\sqrt{\piE (\mbf{p} )}} , 
\end{equation}
where 
$ u_{s, \mbf{x}}^{l}(\mbf{p} ) =  u_{s}^{l}(\mbf{p} ) 
e^{-i \mbf{p} \cdot \mbf{x} }  \; $   and  
   $ \; \tilde{v}^{l}_{s, \mbf{x}} ( \mbf{p} ) = v_{s}^{l}(- \mbf{p} ) e^{-i \mbf{p} \cdot \mbf{x} } $. \\ 
The field operator for electron  is defined by 
\begin{equation}
\psi_{l}(\mbf{x}) = \sum_{s=\pm \half}( b_{s} (g_{s,\mbf{x}}^{l} ) +   d^{\ast}_{s} (h_{s, \mbf{x}}^{l} )).
\end{equation}
We assume the following conditions.  
\begin{quote}
\textbf{(A.2) (Ultraviolet cutoff for the Dirac field) }
\[
\int_{\Rthree} \left| \frac{\chi_{\el} (\mbf{p}) }{
 \sqrt{E_{M}(\mbf{p}) }} \right|^{2} d \mbf{p}  \; <  \; \infty .
\]
\end{quote}
It is  seen that $b_{s} (f) $ and $d_{s} (f) $  are bounded 
 with $ \| b_{s} (f)  \|  = \| d_{s} (f)  \| = \| f \| $.
Then, we can see that 
\begin{equation}
\| \psi_{l} ( \mbf{x} ) \| \leq M_{l}^{\el}, \quad \quad  \quad   l=1,\cdots , 4 ,
\label{boundpsil}
\end{equation}
where $ M_{l}^{\el} = 
\sum_{s=\pm 1/2} 
\left(  \frac{}{} \right. 
 \left\| \frac{\chi_{\el} u_{s}^{ l}}{ \sqrt{(2 \pi)^{3} E_{M}}} 
\right\|
+  \left\| \frac{\chi_{\el} \tilde{v}_{s}^{l }}{ \sqrt{(2 \pi)^{3} E_{M}}} \right\|
 \left. \frac{}{} \right) $.

\subsection{Total Hamiltonian}
The total Hilbert space
 is defined by
\begin{equation}
\ms{F}_{\QED} = \ms{F}_{\el} \tens \ms{F}_{\ph} ,
\end{equation}
 and the decoupled Hamiltonian on  $ \ms{F}_{\QED}$ by 
 \begin{equation}
 H_{0} = H_{\el} \tens I + I \tens H_{\ph} . 
 \end{equation}
 In order to define the interaction, we 
introduce spatial cutoff functions $\chi_{\I}$ and $\chi_{\II} \; $ satisfying the following properties.
\begin{quote}
\textbf{(A.3)}  
\[
\int_{\Rthree} | \chi_{\I} (\mbf{x}) | \dx < \infty , \qquad  
 \int_{\Rthree  \times \Rthree } \frac{| \chi_{\II}
(\mbf{x} )  \; \chi_{\II}(\mbf{y})| }{|\mbf{x} -\mbf{y}|}  
\dx \dy < \infty .
\]
\end{quote}

$\quad $ \\
If $ \chi_{\II} \in L^{6/5} (\Rthree)$,  it follows 
 from  the  Hardy-Littlewood-Sobolev inequality
 (\cite{LiLo}; 4.3 Theorem)  that
\[
\int_{\Rthree \times \Rthree} \frac{| \chi_{\II}(\mbf{x} )  \; \chi_{\II}(\mbf{y})| }{|\mbf{x} -\mbf{y}|}  
\dx \dy \leq \text{const.} \| \chi_{\II} \|_{L^{6/5}}^{2} .
 \]

$\quad $ \\ 
 Let  $\Psi \in \ms{D} (I \otimes H_{\ph}^{1/2} ) $. 
 Then, we can define the functional  $\ell_{\Psi} : 
\ms{F_{\QED}} \to \mbf{C}  $ by
\[
\ell_{\Psi} (\Phi) = \sum_{j} \int_{\Rthree} \chi_{\I}(\mbf{x} ) 
( \psi^{\ast}(\mbf{x}) \alpha^{j} \psi (\mbf{x} ) \otimes A_{j}(\mbf{x} ) \Psi, \, \Phi )_{\ms{F}_{\QED}} \dx  .
\]
 Since  $ \| A_{j}(\mbf{x} ) \Psi \| $ and 
$ \| \psi (\mbf{x} ) \|  $ are uniformly bounded with respect to $\mbf{x}$ by (\ref{boundAj}) and (\ref{boundpsil}),
\begin{equation}
| \ell_{\Psi} (\Phi) | 
 \leq   \left(  L_{\I}  \| (I \otimes H_{\ph}^{1/2} ) \Psi \| + R_{\I} \| \Psi \| \right) \| \Phi \|  \label{11/28.1}
\end{equation}
follows, where
\begin{align}
&L_{\I} = 2 \| \chi_{\I} \|_{L^{1}} \sum_{j, l, l',r } 
| \alpha_{l, \, l'}^{j} | \, 
\, M_{l}^{\el} \, M_{l'}^{\el} \, M_{2 ,j,  \, r}^{\ph} , 
\label{LsubI}  \\
&R_{\I} = \| \chi_{\I} \|_{L^{1}} \sum_{j, l, l',r } 
| \alpha_{l, \, l'}^{j} | \, 
\, M_{l}^{\el}M_{l'}^{\el} \, M_{1 ,j,  \, r}^{\ph} .
\label{RsubI}
\end{align}
Here, we used $\| \chi_{\I} \|_{L^{1}} < \infty $ in  
\textbf{(A.3)}. 
 By the Riesz representation theorem, there exists a unique vector $ \, \Xi_{\Psi}  \in  \ms{F_{\QED}} \, $ 
such that 
\[
 \, \ell_{\Psi}(\Phi)= ( \Xi_{\Psi} , \Phi ) \qquad
 \text{ for all } \quad \Phi \in \ms{F_{\QED}}  .
\]
 Let us define $H'_{\I} \, : \, \ms{F}_{\QED} \; \to  \ms{F}_{\QED} $ by 
\begin{equation}
H'_{\I} : \Psi \longmapsto \Xi_{\Psi} . 
\end{equation}
It is seen from (\ref{11/28.1}) that
\begin{equation}
 \| H'_{\I}   \Psi \| \leq 
  L_{\I}  \| (I \otimes H_{\ph}^{1/2} ) \Psi \| + R_{\I} \| \Psi \| . \label{InteractboundI}
\end{equation}
We may denote $H'_{\I}$ formally by 
\[
H'_{\I} = \int_{\Rthree} \chi_{\I}(\mbf{x} ) 
\psi^{\ast}(\mbf{x} ) \alpha^{j} \psi (\mbf{x} ) \tens
 A_{j}(\mbf{x} ) \,  \dx \, .
\]

$\quad $  \\ 
In the similar as  $H_{\I}'$, let us define the functional 
 $q_{\Psi} : \ms{F_{\QED}} \to \mbf{C}  $ by
 \[
 q_{\Psi} (\Phi ) =
 \int_{\Rthree \times \Rthree} 
\frac{\chi_{\II}(\mbf{x} )  \chi_{\II}(\mbf{y})}{|\mbf{x} - \mbf{y}|}
\left( \psi^{\ast}(\mbf{x} ) \psi (\mbf{x} ) \psi^{\ast}(\mbf{y})\psi (\mbf{y})
 \otimes I  \Psi, \Phi \right)_{\ms{F}_{\QED}} \dx \, \dy .
 \]
 It is seen that by \textbf{(A.3)}
\begin{equation}
 | q_{\Psi} (\Phi ) | 
 \leq \left( M_{\II} \sum_{l, \nu , }
 ( M_{l}^{\el}  M_{\nu}^{\el} )^{2} \right) 
 \| \Psi \|    \|  \Phi \| ,    \label{11/28.2}
\end{equation} 
where  $ M_{\II} :=\int_{\Rthree  \times \Rthree } \frac{| \chi_{\II}
(\mbf{x} )  \; \chi_{\II}(\mbf{y})| }{|\mbf{x} -\mbf{y}|}  
\dx \dy $. 
 Then by the Riesz representation theorem,
 there exists a unique vector $\Upsilon_{\Psi} \in \ms{F_{\QED }}  $ such that  
 \[
  q_{\Psi}(\Phi ) = ( \Upsilon_{\Psi} , \Phi ) \qquad
 \text{ for all } \quad \Phi \in \ms{F_{\QED }} .
\]
 Then we can define $H'_{\II}$  by
\begin{equation}
H'_{\II} : \Psi \longmapsto \Upsilon_{\Psi} . 
\end{equation}
By (\ref{11/28.2}), it is seen that $H_{\II}$ is bounded with 
\begin{equation}
\| H_{\II}' \| \leq 
 M_{\II} \sum_{l, \nu }
 ( M_{l}^{\el}  M_{\nu}^{\el} )^{2} .
 \label{InteractboundII}
\end{equation}
  We may also denote $H'_{\II}$ formally by 
\[
H'_{\II} 
=\int_{\Rthree \times \Rthree} 
\frac{\chi_{\II}(\mbf{x} )  \chi_{\II}(\mbf{y})}{|\mbf{x} - \mbf{y}|}
\psi^{\ast}(\mbf{x} ) \psi (\mbf{x} ) \psi^{\ast}(\mbf{y})\psi (\mbf{y})
 \otimes I   \dx \, \dy .
\]
Now let us define the total Hamiltonian under consideration by
\begin{equation}
H= H_{0} +  H' ( \kappa_{\I} , \kappa_{\II} ),
\end{equation}
where 
\begin{equation}
 \qquad 
   H' ( \kappa_{\I} , \kappa_{\II} )
 = \kappa_{\I} H'_{\I} + \kappa_{\II} H'_{\II} ,  \qquad 
 \kappa_{\I}, \kappa_{\II} \in \mbf{R}   .
 \end{equation}

\begin{lemma} \textbf{(Self-adjointness)} \\
Assume that \textbf{(A.1)}-\textbf{(A.3)} hold. Then, $H$ is self-adjoint on  $\domain{H_{0}}$. 
 Moreover, $H$ is essentially self-adjoint on any core of $H_{0}$ and bounded from below. 
\end{lemma}

\begin{remark}
By the previous lemma, it can be  seen that $H$ is
 essentially self-adjoint on 
\begin{equation}
\ms{D}_{0} := \core ,  \label{11/14.1}
\end{equation}
where $\hat{\tens}$ denotes the algebraic tensor product.
\end{remark}

$\quad$ \\ 
 Next let us consider the spectrum of $H_{\el}$ and    $H_{\ph}$. It is well known that
$ \sigma (H_{\el}) = \{ 0 \} \cup [M , \infty ) $, 
   $ \; \sigma (H_{\ph}) = [ 0 , \infty )$  and 
 $ \sigma_{p} (H_{\el})  = \sigma_{p} (H_{\ph}) = \{ 0 \} $. 
Then, the  spectrum of $ H_{0} = H_{\el} \tens I + I \tens H_{\ph}$ is 
  $  \sigma (H_{0}) = [ 0 , \infty )  \;$
and the point spectrum is $\sigma_{p} (H_{0})  = \{ 0 \} $.
It is also seen that  $ H_{0} \Omega_{0} = 0 $, 
where $ \Omega_{0}= \Omega_{\el} \tens \Omega_{\ph}$.
  Since the ground state energy $0$ of $H_{0}$ is embedded in 
$[0, \infty )$, it is not trivial to see 
that $H$ has the ground state for nonzero $\kappa_{\I} $ and
 $\kappa_{\II}$.

$\quad$ \\ 
To prove the existence of the ground state of $H$, we introduce following assumptions.
\begin{quote}
\textbf{(A.4)} 
It holds that $ \chi_{\ph} \in L^{1}_{loc} (\Rthree )$. 
 
$\quad $ \\ 
\textbf{(A.5)}  It holds that
\[
\int_{\Rthree} \;  | \mbf{x} | \, | \chi_{\I} (\mbf{x} ) |  
\dx  \; <\;  \infty. 
\]

$ \quad $ \\
\textbf{(A.6)} \textbf{(Infrared regularity condition)} \\ 
It holds that
\[ 
\qquad 
\int_{\Rthree} \left|
\frac{ \chi_{\ph} ( \mbf{k}) }{ \sqrt{\omega (\mbf{k})}^{3} }  \right|^{2} \dk 
,\quad  \quad j=1,2,3, \, \,
r=1,2 .
\]
\end{quote}
\begin{theorem}  \label{GROUNDSTATE}
\textbf{(Existence of ground state)} \\
Assume that \textbf{(A.1)}-\textbf{(A.6)} hold.   
   Then for  sufficiently small  $ | \kappa_{I} |$ and $ | \kappa_{II} |$,   $ H$ has a ground state.
\end{theorem}

$\quad$ \\ 
Next let us investigate the multiplicity of the ground states. 
In order to show the uniqueness of the ground state and the existence of the asymptotic field, we make a stronger assumption than \textbf{(A.4)}.
\begin{quote}
\textbf{(A.7)} 
$ \quad $  
There exists a closed set $O_{\ph} \subset \Rthree$
 with the  zero Lebesgue   measure  such that   $\chi_{\ph} \in  C^{\infty} (\Rthree \backslash 
  O_{\ph}  )$ and  
$ \; e_{r}^{j} \in C^{\infty} (\Rthree \backslash 
  O_{\ph}) , \quad j=1,2,3, \, \,r=1,2 $.
\end{quote}
\begin{proposition} \label{UNIQUENESSOFGS}
\textbf{(Uniqueness of ground state)} \\ 
Assume  \textbf{(A.1)}-\textbf{(A.7)}.  
 Then, for sufficiently small $ | \kappa_{I} |$ and $ | \kappa_{II} |$,   \textrm{dimker} $(H-E_{0}(H)) =1$.
\end{proposition}

In addition, we investigate the spectral scattering theory.
Let us assume the following condition.
\begin{quote}
\textbf{(A.8)} 
There exists a closed set $O_{\el} \subset \Rthree$
 with the Lebesgue measure zero 
 such that 
\begin{equation}
 \chi_{el}, \;  \; u_{s} ,  \; \;   v_{s} \in C^{\infty} (\Rthree \backslash O_{\el} ), \qquad s= \pm 1/2 .
\label{11/30.1}
\end{equation} 
\end{quote}
Example :  \\
Let us take the standard representation  
 \[
   \alpha^{j} =  \left( 
\begin{array}{cc}
0 & \sigma^{j} \\
\sigma^{j}  & 0
\end{array}
\right) , 
 \qquad
 \beta = 
 \left( 
\begin{array}{cc}
I & 0 \\
0  & I
\end{array}
\right) ,
\] 
where
 $ \sigma^{1} = \left( 
\begin{array}{cc}
0 & 1 \\
1  & 0
\end{array}
\right)  $,  
 $ \sigma^{2} = \left( 
\begin{array}{cc}
0 & -i \\
i  & 0
\end{array}
\right)
$ and 
$ \sigma^{3} = \left( 
\begin{array}{cc}
1 & 0 \\
0  & -1
\end{array}
\right)  $. 
Then, the angular momentum of the spin is 
 $ \; \mbf{s} =\left( 
\begin{array}{cc}
 \mbf{\sigma} & 0 \\
0  & \mbf{\sigma}
\end{array}
\right)   $. 
Let us set $ O = \{  \mbf{p}  = (p_{1} , p_{2}, p_{3} )  \; \left| \frac{}{} 
 \right. p_{1} \ne 0 \text{ or } p_{2} \ne 0 \}  $ and 
\[ 
\phi_{+} (\mbf{p} ) = \left\{
\begin{array}{cc}
\frac{1}{\sqrt{ 2 | \mbf{p} | (  | \mbf{p} | -p_{3} ) } }
 \begin{pmatrix}
  p_{1} - i p_{2}  \\ p_{3} - | \mbf{p} |  
\end{pmatrix}  \;   &  \mbf{p}  \notin O    ,  \\ 
  \begin{pmatrix} 
  1 \\ 0   \end{pmatrix} 
   & \mbf{p}  \in O   ,
 \end{array} 
 \right.
\quad   
 \phi_{-} (\mbf{p} ) = 
 \left\{
 \begin{array}{cc}
\frac{1}{\sqrt{ 2 | \mbf{p} | (  | \mbf{p} | -p_{3} ) } }
    \begin{pmatrix}
p_{3} - | \mbf{p} |  \\  p_{1} + i p_{2} 
\end{pmatrix}  
  &  \mbf{p}  \notin O    ,  \\ 
  \begin{pmatrix}   0  \\  1 \end{pmatrix}   & 
  \mbf{p}  \in O
  \end{array}
\right.
\]
Let $ \lambda_{\pm} (\mbf{p}) = 
\frac{1}{\sqrt{2}} \sqrt{  1 \pm E_{M} (\mbf{p})^{-1 }}$.   
Then it is seen that  
  \[
 u_{\pm1/2} (\mbf{p}) = 
 \begin{pmatrix} \lambda_{+} (\mbf{p} ) \phi_{\pm} (\mbf{p}) 
 \\  \pm \lambda_{-} (\mbf{p} ) \phi_{\pm} (\mbf{p})   \end{pmatrix} ,  \qquad
 v_{\pm 1/2} (\mbf{p}) = 
 \begin{pmatrix} \mp \lambda_{-} (\mbf{p} ) \phi_{\pm} (\mbf{p}) 
 \\  \pm \lambda_{+} (\mbf{p} ) \phi_{\pm} (\mbf{p})   \end{pmatrix} .
 \]
Then $ u_{\pm s} , \;  u_{\pm s} \; \in 
 C^{\infty} ( \Rthree \backslash   O )  $.

\begin{theorem} \label{asympphoton}
\textbf{(Asymptotic  photon fields)} \\ 
Suppose \textbf{(A.1)}-\textbf{(A.3)},\textbf{(A.6)} and 
 \textbf{(A.7)}. Let 
$ \xi \in \domain{\omega^{-1/2}} $.
Then for $ \Psi \in \domain{H}$, the asymptotic field
\[ 
a_{r, \pm \infty}^{\sharp} (\xi ) \Psi
 := s-\lim_{t \to \pm \infty}
e^{itH} e^{-itH_{0}} (I \otimes a_{r}^{\sharp} (\xi ))e^{itH_{0}} e^{-itH} \Psi  ,
\]
exists. \\ 
\end{theorem}
\begin{theorem} \label{asympDirac}
\textbf{(Asymptotic  Dirac fields)}
  \\ 
Suppose \textbf{(A.1)}-\textbf{(A.3)} and \textbf{(A.8)}.
Let $\eta, \; \zeta \in \eltwo $. 
Then, the asymptotic fields
 \begin{align*}
 & b_{s, \pm \infty}^{\sharp} (\eta ) 
 :=  s-\lim_{t \to \pm \infty }
e^{itH} e^{-itH_{0}}( b_{s}^{\sharp}(\eta ) \otimes I)
e^{itH_{0}} e^{-itH} ,  \\ 
& d_{s, \pm \infty}^{\sharp} (\zeta )  
 :=  s-\lim_{t \to \pm \infty }
e^{itH} e^{-itH_{0}} (d_{s}^{\sharp}(\zeta ) \otimes I )
e^{itH_{0}} e^{-itH}  
 \end{align*}
 exist.
\end{theorem}

By using the asymptotic fields, we can obtain the following theorem.
\begin{theorem} \label{spectrumgap}
 \textbf{(Absence of  spectral gap)} \\ 
Suppose that \textbf{(A.1)-(A.8)} hold. Then $\,  \sigma (H) = [ E_{0}(H), \infty ) $.
\end{theorem}

Finally, we consider the total charge of the ground state. 
The number operators of the electron and the positron
\begin{equation}
N_{+} = \sqzf{{}^{t}(1,1,0,0) }  , \qquad 
N_{-} = \sqzf{{}^{t}(0,0,1,1) } ,
\end{equation}
respectively, and the total charge
\begin{equation}
Q= N_{+} - N_{-} .
\end{equation} 
Since $\psi^{\ast}(\mbf{x})    \psi (\mbf{x}) $ leaves the total charge invariant 
$ [  \psi^{\ast}(\mbf{x})    \psi (\mbf{x}), Q] =0$, it is proven
 in Lemma \ref{12/1.1} that $H$ also leaves the total charge invariant
 $ e^{it Q \tens I } H  e^{- it Q \tens I } = H $. Then, $\ms{F}_{\QED}$ is decomposed with respect to the spectrum of the total charge as
\begin{equation}
\ms{F}_{\QED} = \bigoplus_{z \in \mathbb{Z} } \ms{F}_{z} .
\end{equation}
We  will prove  that the total charge of the ground state is zero.  
 \begin{theorem} \textbf{(Total charge of ground state)} \\ 
 \label{TOTALCHARGEOFGS}
Assume \textbf{(A.1)}-\textbf{(A.6)}.  
   Let  $\Psi_{g}$ be the ground state of $H$. 
Then for sufficiently small $ | \kappa_{I} |$ and 
  $ | \kappa_{II} |$,  $ \; \Psi_{g} \in \ms{F}_{0}$.
\end{theorem}

\section{Ground States}
\subsection{Self-adjointness}

It is noted that, by the spectral decomposition theorem,
for all $\epsilon > 0$, there exists a positive number $c_{\epsilon} \, > \, 0$ such that for all $\Psi \in \ms{D}(H_{\ph})$,
\begin{equation}
\| H_{\ph}^{1/2} \Psi \| \leq
\epsilon \| H_{\ph} \Psi \| + c_{\epsilon} \| \psi \| .
\label{halfepsilon}
\end{equation}
  \textbf{(Proof of Lemma1.1)} \\
By (\ref{halfepsilon}) and (\ref{InteractboundI}), we see that for $\Psi \in \ms{D} (H_{0})$, 
\begin{equation}
\| H'_{\I} \Psi \| \leq \epsilon L_{\I} \| H_{0} \Psi \|  + 
(c_{\epsilon} L_{\I} +R_{\I})  
\| \Psi \| . \label{8/30.1}
\end{equation}
From (\ref{InteractboundII}) and (\ref{8/30.1}), it follows that for $\Psi \in \ms{D} (H_{0})$,
\begin{equation}
\|  H'(\kappa_{\I}, \kappa_{\II} )   \Psi  \| \leq \epsilon | \kappa_{\I} |  L_{\I} \| H_{0} \Psi \| + 
(| \kappa_{\I} | (c_{\epsilon} L_{\I} +R_{\I} )
+ \kappa_{\II} \| H_{\II}' \| )   \| \Psi \| . 
\label{Interactionepsilon}
\end{equation}
Let us take $\epsilon >0 $ such as $ \epsilon | \kappa_{I} | L_{I}< 1 $. Then the Kato-Rellich theorem reveals that  
$ H $ is self-adjoint on  $\ms{D} (H_{0})$, essentially self-adjoint on any core of $H_{0}$, and bounded from below. 
$\blacksquare$

\subsection{Existence of Ground State}
To prove the existence of a ground state of $H$, we introduce some 
 Hamiltonians  approximating $H$. 
For $\; m>0$, let 
$ \omega_{m} (\mbf{k} ) = \omega ( \mbf{k} ) + m $ and $ H_{\ph}(m) = \sqzb{\omega_{m}} $.  
Let 
\begin{equation}
H(m) = H_{0} (m) + H'( \kappa_{\I} , \kappa_{\II}) ,
\end{equation}
where $  H_{0} (m) 
=  H_{\el} \otimes I + I \otimes H_{\ph} (m) $.

$\quad$ \\ 
In order to prove the existence of a ground state of $H(m)$, we apply the momentum lattice approximation (e.g. 
 \cite{AH97}, \cite{BFS99}, \cite{BDG04}, \cite{DiGu03}, 
 \cite{GJ70} , \cite{Hiro}, \cite{Hi01} ).
For $V>0$ and $ L>0$, we set
\begin{align*}
&\Gamma_{V}= \frac{2\pi}{V} \mbf{Z}^{3}
=\{ \mbf{q} = (q_{1},q_{2},q_{3}) | q_{j} = \frac{2\pi}{V} n_{j}  ,\; \;  n_{j} \in \mbf{Z} , \;  j=1,2,3     \} ,  \\ 
&\Gamma_{V,L} = \{  \mbf{q} = (q_{1},q_{2},q_{3})  \in  \Gamma_{V}
\; |  \; \;  |q_{j}| +  \frac{\pi}{V} \leq L, \; j=1,2,3 \}  ,
\end{align*}
and $ \ms{F}_{\ph, \, V } = \ms{F}_{\bos} (\ell^{2} ( \Gamma_{V} )  \oplus \ell^{2} ( \Gamma_{V} )  )$.  
We can identify $ \; \ms{F}_{\ph, \, V } \; $ with a closed  subspace of $\;  \ms{F}_{\ph} $. 
For a lattice point $ \; \mbf{q} \in \Gamma_{V} $, we set 
$C_{\mbf{q} ,V } = \prod_{j=1}^{3} [ q_{j}-\frac{\pi}{V}, q_{j}+\frac{\pi}{V} )  \subset \Rthree$. 

$\quad$ \\
Let
\[
\omega_{m,V} (\mbf{k} ) = \sum_{\mbf{q} \in \Gamma_{V}} \omega_{m} (\mbf{q} ) 
\chi_{C_{\mbf{q},V} }(\mbf{k} ), \qquad
(f_{r, \mbf{x} }^{j} )^{L,V} (\mbf{k} ) = \sum_{\mbf{q} \in \Gamma_{V,L}} f_{r, \mbf{x} }^{j} 
 (\mbf{q} ) \chi_{ C_{\bf{q},V} }(\mbf{k} ) ,
\]
   where $\chi_{ C_{\bf{q},V} }$  is the characteristic function on $C_{\bf{q},V}  $.  In addition let us set
 $ \;  (f_{r, \mbf{x} }^{j} )^{L} (\mbf{k})  $ $ = \chi_{L} (\mbf{k} ) 
(f_{r, \mbf{x} }^{j} )(\mbf{k} ) \; $ for  $ \; \chi_{L} (\mbf{k} ) = \chi_{[-L,L]}(k_{1} ) \chi_{[-L,L]}(k_{2} ) \chi_{[-L,L]}(k_{3} ) $. 
Let $\ms{F}_{V} = \ms{F}_{\el} \otimes \ms{F}_{\ph ,V} $. 
 We introduce the operators  
\[
H_{0,V} (m) = H_{\el} \otimes I + 
I \otimes H_{\ph,V} (m)  ,
\]
where $ H_{\ph ,V} (m) = \sqzb{\omega_{m,V}} $, and
\begin{align*}
& H_{\I,L,V}' = \sum_{j} \int_{\Rthree} \chi_{\I}(\mbf{x} ) 
 \left( \psi^{\ast}(\mbf{x}) \alpha^{j} \psi (\mbf{x} ) \otimes  A_{j}^{L,V} (\mbf{x} ) \right) \dx , \\ 
 & H_{\I,L}' = \sum_{j} \int_{\Rthree} \chi_{\I}(\mbf{x} ) 
 \left( \psi^{\ast}(\mbf{x}) \alpha^{j} \psi (\mbf{x} ) \otimes  A_{j}^{L} (\mbf{x} ) \right) \dx , 
\end{align*}
where
$  A_{j}^{L,V} (\mbf{x} ) = \sum\limits_{r=1,2} 
 \left( a_{r} ((f_{r, \mbf{x} }^{j} )^{L,V} ) + 
 a_{r}^{\ast}((f_{r, \mbf{x} }^{j} )^{L,V}) \right) , \quad   
 A_{j}^{L} (\mbf{x} ) = \sum\limits_{r=1,2} 
 \left( a_{r} ((f_{r, \mbf{x} }^{j} )^{L} ) + 
 a_{r}^{\ast}((f_{r, \mbf{x} }^{j} )^{L} )\right)  $. \\
Let us set
\begin{align}
&H_{L,V} (m) = H_{0,V} (m) \;
 + \kappa_{\I} H'_{\I, L,V} + \kappa_{\II}  H'_{\II}  \\
&H_{L} (m)= H_{0} (m)\; 
 + \kappa_{\I} H'_{\I, L} + \kappa_{\II}  H'_{\II}  .
\end{align}
In similar fashion to the proof of Lemma 1.1, it can be proven that
$H_{L,V}(m)$ and $H_{L}(m)$ are self-adjoint, and essentially self-adjoint on any core of $H_{0,V} $ and $H_{0}(m)$,  respectively.   
In particular $ H_{L,V} (m)$ is essentially self-adjoint on 
$ \ms{D}_{0,V}^{m} :=$ $ \coremV $,
and $ H_{L} (m)$ on $ \ms{D}_{0}^{m}:= 
\ms{F}^{\textrm{fin}}_{\textrm{el}} (\ms{D}(E_{M})) $ 
 $ \hat{\otimes} $ $ \ms{F}_{\textrm{ph}}^{\textrm{fin}}  (\ms{D}(\omega_{m})) $.
 
.    
\begin{lemma} \label{12/15.5}
Assume  \textbf{(A.1)}-\textbf{(A.3)}. Then 
\begin{equation}
E_{0}(H) \leq | \kappa_{II} | \| H_{II}' \| 
\label{bondestimate} 
\end{equation} 
 holds, where $ E_{0} (H) = \inf \sigma (H)$.
\end{lemma}
\textbf{(Proof)} Let $\; \Psi_{\el} \in \ms{D} (H_{\el})\; $  and $\; \| \Psi_{\el} \| =1 $.
For $\Psi = \Psi_{\el} \otimes \Omega_{\ph} $,  
\[
( \Psi , H \Psi ) = ( \Psi_{\el} , H_{\el} \Psi_{\el} )
+ \kappa_{\II} ( \Psi_{\el} , H_{II}' \Psi_{\el} )  \;
\leq  \; ( \Psi_{\el} , H_{\el} \Psi_{\el} ) + |
 \kappa_{\II} | \; \| H_{\II}' \| .
\]
Here we used that  $\; ( \Omega_{\ph}, H_{\ph} \Omega_{\ph}  ) = 0 \; $  and $ \; ( \Omega_{\ph}, A_{j}(\mbf{x} ) \Omega_{\ph}  ) = 0  $. 
Then, $ E_{0} (H)  \leq \; ( \Psi_{\el} , H_{\el} \Psi_{\el} ) + | \kappa_{\II} | \; \| H_{\II}' \|  \; $, and  hence $ E_{0}(H) \leq  E_{0}(H_{\el}) +  | \kappa_{\II} | \; \| H_{\II}' \|  \;  =   | \kappa_{\II} | \; \| H_{\II}' \| $. 
$\blacksquare$

\begin{lemma}   \label{GS-HLV}
Assume \textbf{(A.1)}-\textbf{(A.3)}.  Then \\  
\textbf{(1)}$H_{L,V}(m)$ is reduced by $\ms{F}_{V} $, \\ 
\textbf{(2)}For sufficiently small $m$, $  |\kappa_{\I} | $
 and    $ |\kappa_{\II} | $, 
 $H_{L,V} (m)$ has a purely discrete spectrum in $ [ E_{0}( H_{L,V} (m) ), $ $E_{0}( H_{L,V}(m)  )+ m ) $.
\end{lemma}
\textbf{(Proof)}
\textbf{(1)}
Let $\Psi = \Psi_{\el}\tens \Psi_{\ph} \in \ms{D}_{0,V}^{m}$ with $ 
\Psi_{\ph}= a_{r_{1}}^{\ast} (f_{1}) \cdots  a_{r_{n}}^{\ast} (f_{n})
\Omega_{\ph}, $ $\; f_{j} \in \domain{ \omega_{m,V}} , \, j=1,\cdots ,n  $.
Let $\; q_{V} : \eltwo \to \ell^{2}(\Gamma_{V}) \; $ and
 $\; Q_{V} :\ms{F}_{\ph} \to \ms{F}_{\ph,V} \;$ be the orthogonal projections.
It is seen that 
\begin{equation}
H_{0,V}(m) (I \otimes Q_{V}) \Psi = (I \otimes Q_{V}) H_{0,V}(m) \Psi  .  \label{reduce0}
\end{equation}
 Since $ q_{V} \chi_{C_{\mbf{q} ,V}}=  \chi_{C_{\mbf{q} ,V}}$, 
 we also see that, by the definition of $A_{j}^{L,V} (\mbf{x} ) $,
\begin{equation}
Q_{V} A_{j}^{L,V}(\mbf{x} )  \Psi_{\ph} =A_{j}^{L,V}(\mbf{x} )  Q_{V} \Psi_{\ph} ,
 \end{equation}
 follows. Then, for $ \Phi \in \ms{F} $,
\begin{align*}
( \Phi, (I\otimes Q_{V})   H_{\I ,L,V}' \Psi )  
&= \sum_{j} \int_{\Rthree} \chi_{\I}(\mbf{x} ) 
( \Phi, \psi^{\ast}(\mbf{x}) \alpha^{j} \psi (\mbf{x} ) \otimes (Q_{V} A_{j}^{L,V} (\mbf{x} ) )\Psi ) \dx \\
 &= \sum_{j} \int_{\Rthree }\chi_{\I}(\mbf{x} ) 
( \Phi, \psi^{\ast}(\mbf{x} ) \alpha^{j} \psi (\mbf{x} ) \otimes ( A_{j}^{L,V} (\mbf{x} ) Q_{V} )\Psi ) \dx \\
&= \; ( \Phi, H_{\I,L,V}'(I \otimes Q_{V})  \Psi ).
\end{align*}
Hence, we have $ (I\otimes Q_{V} )H_{\I ,L,V}' \Psi 
= H_{\I ,L,V}' (I \otimes Q_{V} )\Psi \; $ for 
$ \; \Psi \in  \ms{D}_{0,V}^{m}$. It is trivial to see  that  \\  
$ \;(I\otimes Q_{V} )H_{\II}' \Psi 
= H_{\II}' (I \otimes Q_{V} )\Psi $ for $ \Psi \in  \ms{D}_{0,V}^{m}$. 
Then 
\begin{equation}
(I\otimes Q_{V} )H_{L,V}(m)  = H_{L,V}(m) 
(I \otimes Q_{V} )  ,  \label{11/14.2}
\end{equation}
on $ \; \ms{D}_{0,V}^{m}$.
Since $ \; \ms{D}_{0,V}^{m} \; $ is a core of $H_{L,V}(m)$, we can extend (\ref{11/14.2}) for all 
$ \Psi \in \ms{D} (H_{0,V} (m))$. Therefore, $H_{L,V}(m)$ is reduced by $\ms{F}_{V}$.

$\quad $ \\
\textbf{(2)} $\; $  By (\ref{InteractboundI}), there exist $c_{m,L,V} >0\;$ and  $\;d_{m,L,V} >0 \; $  
such that  
\begin{align*}
| ( \Psi, H_{L,V}(m) \Psi ) - ( \Psi, H_{0,V}(m) ) \Psi ) |
\leq &\kappa_{\I} c_{m,L,V}  ( \Psi, I \otimes H_{\ph,V} (m)
\Psi )   \\
 &\qquad +  ( \kappa_{\I} d_{m,L,V}+ \kappa_{\II} \|H_{\II}' \| ) \| \Psi \|^{2}  .
\end{align*}
 Therefore, it can be seen that 
\begin{align}
&H_{\el} \otimes I + (1-\kappa_{\I} c_{m,L,V})I \otimes H_{\ph ,V} (m)-  
( \kappa_{\I} d_{m,L,V}+ \kappa_{\II} \| H_{\II}' \| ) \leq H_{L,V} (m),
\label{leqHsubLV} \\
&H_{L,V}(m) \leq H_{\el} \otimes I + (1+\kappa_{\I}
 c_{m,L,V})I \otimes H_{\ph,V}(m) + ( \kappa_{\I}
 d_{m, L,V}+ \kappa_{\II} \| H_{\II}' \| )  ,
 \label{geqHsubLV}
\end{align}
where $A \leq B$ denotes that 
$\; (f,Af) \leq (f,Bf) \;$ for $ \; f \in 
  \domain{A} \cap \domain{B}$. 
Let 
\[
X_{L,V}(m) := H_{L,V} (m)- E_{0}(H_{L,V} (m)) -m  .
\]
We shall show that $X_{L,V}(m)$ has a purely discrete spectrum
in $[-m, 0 ) $, and hence $H_{L,V}(m)$ has a purely discrete spectrum in $ [ E_{0}(H_{L,V} (m)), \, E_{0}(H_{L,V}(m))+m )$.
 By (\ref{bondestimate}) and (\ref{leqHsubLV}), we see that  
\begin{align}
& X_{L,V}(m)   \notag \\ 
&\geq  H_{\el} \otimes I + I \otimes 
\{    (1-\kappa_{\I} c_{m,L,V} )    H_{\ph ,V} (m)
-(  \kappa_{\I} d_{m,L,V}+ \kappa_{\II} \| H_{\II}' \|  
+ E_{0}(H_{m,L,V}) + m ) \}  \notag  \\
& \geq  H_{\el} \otimes I + I \otimes 
\{    (1-\kappa_{\I} c_{m,L,V} )    H_{\ph ,V} (m)
-(  \kappa_{\I} d_{m,L,V}+ 2 \kappa_{\II} \| H_{\II}' \|  + m ) \}  . 
\label{XsubLVone}
\end{align}
It is noted that $I= E_{H_{\el}} ([0,M) ) + 
E_{H_{\el}} ([M,\infty )) $ and $H_{\el} \geq M E_{H_{\el}} ([M,\infty ))$, where $ E_{H_{\el}} $ is the spectral projection of $H_{\el}$. 
Let $\kappa_{\I}$, $\kappa_{\II}$ and 
$m$ be small such that $ \; -(\kappa_{\I} d_{m,L,V}+ 2 \kappa_{\II}  \| H_{\II}' \| + m )+M >0 $. Then we have
$ X_{L,V} (m) \geq S_{L,V} (m)$, 
where
\[
S_{L,V} (m) = E_{H_{\el}} ([0,M) ) \tens \{  (1-\kappa_{\I} c_{m,L,V} )    H_{\ph ,V} (m)-(  \kappa_{\I} d_{m,L,V}+ 2 \kappa_{\II} \| H_{\II}' \| + m ) \} .
\]
Let
 $\; \{ e_{n}^{+} \}_{n=0}^{ N_{+}} \; $ and    
$ \; \{ e_{n}^{-} \}_{n=0}^{ N_{-}} $,
 $ N_{\pm} \leq \infty $,
 be complete orthonormal systems of $\ms{F}_{+} :=E_{X_{L,V}(m) } ([0.\infty ))\ms{F_{\QED}} $ and 
$\ms{F}_{-} :=E_{X_{L,V}(m)}((-\infty,0]) \ms{F_{\QED}} \; $,  respectively. For a self-adjoint operator $X$, we set
\[
X^{+} = E_{X}([\, 0 , \infty )) X E_{X}([\, 0 , \infty )) , \quad 
X^{-} = E_{X}((- \infty , 0 ]) X E_{X}(( -\infty , 0] )  ,
\]
where $ E_{X} $ is the spectral projection of $X$.   
Then we have
\begin{equation}
0\geq \text{Tr} \; X_{L,V}(m){ \restr_{\ms{F}_{-}} } 
\geq \sum_{n=1}^{N_{-}}  ( e_{n}^{-} , S_{L,V}(m) e_{n}^{-})  \;
\geq \sum_{n=1}^{N_{-}}  ( e_{n}^{-} , S_{L,V}(m)^{-} e_{n}^{-} ) \, . \label{zerogeqtrace}
\end{equation}
Here we used  $ S_{L,V} (m)= S_{L,V}(m)^{+} + S_{L,V}(m)^{-} $ and 
$ S_{L,V}(m)^{+} \geq 0$.
Since $ E_{H_{el}} \geq 0 \,$, we see that
\[
S_{L,V}(m)^{-} = E_{H_{\el}} ([0,M) ) \tens \left(  (1-\kappa_{\I} c_{m,L,V} )    H_{\ph ,V}(m) -(  \kappa_{\I} d_{m,L,V}+ 2 \kappa_{\II} \| H_{\II}' \| + m ) \right)^{-} .
\]
Then it follows from (\ref{zerogeqtrace})  that 
\begin{align*}
&| \text{Tr}(H_{L,V}(m) -E_{0}(H_{L,V}(m)) -m )|  \\
& \leq  \text{Tr} \, E_{H_{\el}}([0,M)) \times 
|\text{Tr} \, \left( (1-\kappa c_{m,L,V})H_{\ph ,V}(m) 
-  ( \kappa_{\I} d_{m,L,V}+ 2 \kappa_{\II} \| H_{\II}' \| + m ) 
  \right)^{-} | < \infty .
\end{align*}
Hence  $X_{L,V}(m)$ has a purely discrete spectrum
in $[-m, 0 ) $.

$\quad$ \\ 
Let $\,  M_{V} = \ell^{2}(\Gamma_{V})\oplus  \ell^{2}(\Gamma_{V})  $.  We can decompose $\ms{F}_{\ph}$ as
$ \ms{F}_{\ph} 
\simeq \oplus_{n=0}^{\infty} \left( \ms{F}_{\ph ,V} \otimes ( \otimes_{s}^{n} \ms{M}_{V}^{\bot} )\right) $,
and hence $ 
\ms{F_{\QED}} \simeq \ms{F}_{V} \otimes \ms{F}_{V}^{\bot}$,  
where
$ \ms{F}_{V}^{\bot} = \oplus_{n=1}^{\infty} \left( \ms{F}_{V} \otimes ( \otimes_{s}^{n} \ms{M}_{V}^{\bot} )\right) $.
For $n \geq 1$,
\[
H_{L,V \restr_{ \ms{F}_{V} \otimes ( \otimes_{s}^{n} 
\ms{M}_{V}^{\bot} )}} \simeq H_{L,V \restr_{\ms{F}_{V}}} \otimes I_{ \restr_{\otimes_{s}^{n} \ms{M}_{V}^{\bot}} } 
+ I_{\restr_{\ms{F}_{V}}} \otimes d\Gamma_{\bos}  (\omega_{V})_{\restr_{\otimes_{s}^{n} \ms{M}_{V}^{\bot} } } 
\geq E_{0} ( H_{L,V}  ) + nm.
\]
Hence, $H_{L,V \restr_{\ms{F}_{V}^{\bot}}} \geq E_{H_{0,V}} +m $. $\blacksquare$

\begin{lemma} \label{Helboundness}
Assume \textbf{(A.1)}-\textbf{(A.5)}.
For sufficiently large $L$, there exist constants
 $a_{1} (m) >0 $ and $b_{1} (m)> 0 $ independent of $L$ such that
\begin{equation}
\| H_{0} (m) \Psi \| \leq a_{1}(m) \| H_{L}(m) \Psi \| 
+ b_{1}(m) \| \Psi \|, \quad \quad \Psi \in \ms{D}(H_{0}(m)) .
\end{equation}
\end{lemma}
\textbf{(Proof)}  
Let $\Psi  \; \in \ms{D} (H_{0})$. It is seen that
\begin{align*}
 \| H_{0} (m)\Psi \| &= \| H_{L} (m) \Psi
 - (\kappa_{\I} H_{\I ,L} ' (m)
+ \kappa_{\I} H_{\II}'  ) \Psi \|   \\
 & \leq     \| H_{L} (m) \Psi \| +  |\kappa_{\I}| 
\|H_{\I ,L} ' \Psi \| +
 |\kappa_{\II}| \| H_{\II}' \|  \| \Psi \| .
\end{align*}
Note  that 
\[
\| A_{j}^{L}(\mbf{x}) \Psi \| \leq
2 \| \frac{(\chi_{\ph} e_{r, \mbf{x} }^{j})^{L}}{\sqrt{2 (2\pi)^{3} \omega_{m}  \omega}} \| \| H_{\ph}(m)^{1/2} \Psi \| 
+ \| \frac{(\chi_{\ph} e_{r, \mbf{x} }^{j})^{L}}{\sqrt{2 (2\pi)^{3}\omega_{m}}} \| \| \Psi \| ,
\]
and  $ \;  \lim_{L \to \infty } \| \frac{(\chi_{\ph} e_{r, \mbf{x} }^{j})^{L}}{\sqrt{\omega_{m} \omega  }} \| 
= \| \frac{\chi_{\ph} e_{r, \mbf{x} }^{j}}{\sqrt{\omega_{m} \omega }} \| \; $  and 
$ \;  \lim_{L \to \infty } \| \frac{(\chi_{\ph} e_{r, \mbf{x} }^{j})^{L}}{\sqrt{\omega_{m} }} \| 
= \| \frac{\chi_{\ph} e_{r, \mbf{x} }^{j}}{\sqrt{\omega_{m}  }} \| $.
Hence, for sufficiently large $L$, $\, \| A_{j}^{L}(\mbf{x}) \Psi \| $ is bounded uniformly in $L$. By (\ref{halfepsilon}), it is seen that for all 
 $\epsilon> 0$, there exists a constant $\tilde{c}_{\epsilon}$ such that $ \| H_{I,L}' \Psi \| \leq \epsilon \| H_{0}(m) \Psi \|
+ \tilde{c}_{\epsilon} \| \Psi \| $.  
Hence, we have
\[
\| H_{0} (m) \Psi \| \leq  \frac{1}{1-\epsilon} \| H_{L}(m) \Psi \|  
+  \frac{\tilde{c} +\kappa_{\II} \| H_{II}' \| }{1-\epsilon}
\| \Psi \| , 
\]
and the proof is completed. $\blacksquare$

\begin{lemma} 
Assume \textbf{(A.1)}-\textbf{(A.5)}.
\label{resolventconvergence}
For all $ z \in \mbf{C} \setminus \mbf{R} $, it follows that
\begin{align}
&\lim_{V \to \infty} \| ( H_{L,V} (m) -z )^{-1} -(H_{L} (m)-z )^{-1} \| = 0, 
 \label{rconv1}  \\
&\lim_{L \to \infty} \| ( H_{L} (m)-z )^{-1} -(H(m) -z )^{-1} \| = 0. 
 \label{rconv2}
\end{align}
\end{lemma}
\textbf{(Proof)} $\quad $ \\
 We see that
\begin{align}
( H_{L,V} (m)-z )^{-1} -(H_{L}(m) -z )^{-1}  
 &=   (H_{L,V} (m)-z )^{-1}(I \otimes
  (H_{\ph} (m) -H_{\ph ,V} (m) ) ) 
 (H_{L} (m)  -z )^{-1}  \notag  \\
& \qquad \; +\kappa_{\I} ( H_{L,V} (m)-z )^{-1} 
(H_{\I ,L}' - H_{\I ,L,V}'  ) (H_{L}(m) -z )^{-1} .
\label{AH3.1.1}
\end{align}
Let $ \; C_{V} (m) = \sqrt{3} \left( \frac{\pi}{V} \right)^{3}( \frac{1}{2m} +1 )$. 
Then, it is seen (\cite{AH97} ; Lemma 3.1) that  for $\Psi \in \ms{D} (H_{\ph}(m)) $,
\[ \| ( H_{\ph}(m) -H_{\ph ,V}(m) ) \Psi \| \leq
\frac{2C_{V}(m)}{1-C_{V}(m)} \| H_{\ph}(m) \Psi \| ,
\] 
hence, we obtain
\[  
 \| (I \otimes   (H_{\ph}(m) -H_{\ph ,V} (m) ) )  (H_{L}(m) -z )^{-1} \|  
\leq \frac{2 C_{V}(m)} {(1-C_{V}(m))} \| (I \otimes H_{\ph} (m)) (H_{L}(m)-z)^{-1} \|
\to 0  ,
\] 
 as $V \to  \infty $.
The second term on the right-hand side of (\ref{AH3.1.1}) can be estimated as  
\begin{align*}
&\quad \| (H_{\I ,L}' - H_{\I ,L,V}' )
( H_{L}(m)-z)^{-1} \Psi \| \\
&\leq \sum_{j,l,l'} \,| \alpha_{l, l'}^{j}\, | \; 
 M_{l}^{\el}M_{l'}^{\el}
 \int_{\Rthree} | \chi_{\I} (\mbf{x} ) | 
\| I \otimes \left(   A_{j}^{L}(\mbf{x} ) - A_{j}^{L,V}
(\mbf{x} )  \right) (H_{L}(m)-z)^{-1}  \Psi \|  \dx \, .
\end{align*}
By (\ref{boundar}) and (\ref{boundadr})
we see that for $\Xi \in \domain{H_{\ph}(m)^{1/2}}$
\begin{align*}
\| (   A_{j}^{L}(\mbf{x} ) - A_{j}^{L,V} 
(\mbf{x} )  ) \Xi \| 
 &\leq \frac{1}{\sqrt{2( 2 \pi )^{3}}} \sum_{r} 
\left( \|  \frac{2}{\sqrt{\omega_{m}}} ( \frac{(\chi_{\ph} e_{r, \mbf{x} }^{j})^{L}}{\sqrt{\omega}} - \frac{(\chi_{\ph}e_{r, \mbf{x} }^{j})^{L,V}}{\sqrt{\omega_{V}}}   ) H_{\ph}(m)^{1/2} \Xi  \|  \right. \\
& \left. \qquad \qquad \qquad \qquad + \|   ( \frac{(\chi_{\ph} e_{r, \mbf{x} }^{j})^{L}}{\sqrt{\omega}} - \frac{(\chi_{\ph} e_{r, \mbf{x} }^{j})^{L,V}}{\sqrt{\omega_{V}}}   ) \Xi \| \right) .
\end{align*}
Hence, in order to prove 
$  \;  \lim_{V \to \infty } \| (H_{\I ,L}' - 
H_{\I ,L,V}' )( H_{L}(m)-z)^{-1} \| = 0 $, 
 it is enough to show that 
\begin{equation}
\lim_{V \to \infty} \int_{\Rthree} | \chi_{\I} (\mbf{x} ) | \, \| \frac{(\chi_{\ph} e_{r, \mbf{x} }^{j})^{L}}{\sqrt{\omega}} - \frac{(\chi_{\ph} e_{r, \mbf{x} }^{j})^{L,V}}{\sqrt{\omega_{V}}} \|
\dx \; = 0  . \label{Lebconv}
\end{equation}
It is seen that,
\begin{align*}
\| \frac{(\chi_{\ph} e_{r, \mbf{x} }^{j})^{L}}{
\sqrt{\omega}} - \frac{(\chi_{\ph} e_{r, \mbf{x} }^{j})^{L,V}}{\sqrt{\omega_{V}}} \|^{2} 
& \leq 2 \int_{I_{L}} \left|
\frac{\chi_{\ph}(\mbf{k}) e_{r}^{j} (\mbf{k})}{\sqrt{
\omega (\mbf{k})}} 
- \sum_{\mbf{q} \in \Gamma_{V,L}}
\frac{\chi_{\ph}(\mbf{q}) e_{r}^{j}(\mbf{q})}{\sqrt{
\omega (\mbf{q})}} \chi_{C_{\mbf{q},V}} (\mbf{k})
\right|^{2}  \dk  \\ 
& \quad + 2 \int_{I_{L}} \left|
\sum_{\mbf{q} \in \Gamma_{V,L}}
 \frac{\chi_{\ph}(\mbf{q}) e_{r}^{j} (\mbf{q})}{\sqrt{
\omega (\mbf{q})}} 
( e^{i \mbf{k} \cdot \mbf{x}} -e^{i \mbf{q} \cdot \mbf{x} } )  \chi_{C_{\mbf{q},V}} (\mbf{k})  \right|^{2}     \dk   .
\end{align*}
By the inequality $|e^{i \mbf{k} \cdot \mbf{x}} -
 e^{i \mbf{q} \cdot \mbf{x}} | \leq   | \mbf{x}| \, | \mbf{k} -\mbf{q} | $,
we obtain
 \[
 \| \frac{(\chi_{\ph} e_{r, \mbf{x} }^{j})^{L}}{\sqrt{
\omega}} - \frac{(\chi_{\ph} e_{r, \mbf{x} }^{j})^{L,V}}{\sqrt{\omega_{V}}} \|^{2}
 \leq  X_{L,V}^{j} 
+ \left| \mbf{x} \right|^{2} Y_{L,V}^{j} ,  
\]
where
\begin{align*}
&X_{L,V} = 2 \int_{I_{L}} \left|
\frac{\chi_{\ph}(\mbf{k}) e_{r}^{j} (\mbf{k})}{\sqrt{
\omega (\mbf{k})}} 
- \sum_{\mbf{q} \in \Gamma_{V,L}}
\frac{\chi_{\ph}(\mbf{q}) e_{r}^{j}(\mbf{q})}{\sqrt{
\omega (\mbf{q})}} \chi_{C_{\mbf{q},V}} (\mbf{k})
\right|^{2}  \dk  , \\ 
& Y_{L,V}^{j}= 2 \int_{I_{L}} \left(
\sum_{\mbf{q} \in \Gamma_{V,L}}
 \frac{| \chi_{\ph}(\mbf{q}) e_{r}^{j} (\mbf{q})|}{\sqrt{\omega (\mbf{q})}} 
 | \mbf{k} -\mbf{q} |
 \chi_{C_{\mbf{q},V}} (\mbf{k})  \right)^{2}     \dk   .
\end{align*}
Hence, we have
\[
 \int_{\Rthree} | \chi_{\I} (\mbf{x} ) | \, 
\| \frac{(\chi_{\ph} e_{r, \mbf{x} }^{j})^{L}}{\sqrt{
\omega}} - \frac{(\chi_{\ph} e_{r, \mbf{x} }^{j})^{L,V}}{\sqrt{\omega_{V}}} \| \dx \leq
\;  \| \chi_{\I} \|_{L^{1}}  \sqrt{X_{V}^{L,j}} +
\int_{\Rthree} | \mbf{x} | \chi_{\I} (\mbf{x} ) | \dx 
\sqrt{Y_{V}^{L,j} } .
\] 
Here, we used the assumption 
$\; \int_{\Rthree} | \mbf{x} | \chi_{\I} (\mbf{x} ) | \dx < \infty $ of \textbf{(A.6)}. 
Then, by the Lebesgue dominated convergence, we see that
 $ \; X_{L,V}^{j} \to 0 \; $  and $\; Y_{L,V}^{j} \to 0 \; $ as $ \; V \to \infty $. Then (\ref{Lebconv}) is obtained, and hence (\ref{rconv1}) follows.
We can prove (\ref{rconv2}) similarly to (\ref{rconv1}) by using Lemma \ref{Helboundness}.
$\blacksquare$

\begin{proposition} \label{proofmassive}
$\quad $ \\ 
Assume \textbf{(A.1)}- \textbf{(A.5)}. Then
$H(m)$ has a purely discrete spectrum in $[ E_{0}(H(m)) , \,
 E_{0}  (  H(m) )  + m )$. In particular $H(m)$ has a ground state.  
\end{proposition}
\textbf{(Proof)}
By Lemma \ref{GS-HLV}, $H_{L,V}(m)$ has a purely discrete spectrum in $[ E_{0}(H_{L,V}(m)),$ $E_{0}(H_{L,V}(m))+ m )$. In addition, $H_{L,V}(m)$ converges to $H_{L}(m)$ in the norm resolvent sense as $V \; \to \; \infty$ 
 by Lemma \ref{resolventconvergence}. Hence, by 
the general theorem (\cite{SH72} ; Lemma 4.6),
  $\, H_{L}(m)$ has a purely discrete spectrum in  $[ E_{0}(H_{L}(m)) , E_{0}(H_{L}(m))+ m )$. It is also seen that
 $H_{L}(m)$ converges to $H$ in the norm resolvent sense
as $L \;\to \;\infty$ by Lemma \ref{resolventconvergence}. Hence, $H(m)$ has a purely discrete spectrum in  $[ E_{0}(H(m)) , E_{0} (H(m))+ m )$ by the same  theorem (\cite{SH72} ; Lemma 4.6). $\blacksquare $

$\quad$ \\ 
By proposition \ref{proofmassive} 
 $H(m)$ has a ground state $\Psi_{m}$:
\[
H(m) \Psi_{m} = E_{0} (H(m)) \Psi_{m} .
\]

$\quad$ \\
The number operator of 
$\ms{F}_{\bos}( L^{2}(\Rthree ; \mbf{C}^{2}) )$ is defined by
\begin{equation}
N_{\ph} = \sqzb{I} .
\end{equation}

\begin{lemma} \label{AH-p482}
Suppose that \textbf{(A.1)} -\textbf{(A.6)}. Then 
\[
\| (I \otimes N_{\ph}^{1/2} ) \Psi_{m} \| \leq
\ |\kappa_{\I} |  \sum_{j,l,l',r}  \nu^{j,l,l'}
\left\|   \frac{\chi_{\ph} e^{j}_{r}}{\sqrt{2(2\pi )^{3} \omega_{m}^{2} \omega }} \right\|  \| \Psi_{m} \| , 
\]
where
$ \;  \nu^{j,l,l'}= \| \chi_{\I} \|_{L^{1}}  | \alpha_{l,l'}^{j} | M_{l}^{\el}M_{l'}^{\el} .$
\end{lemma}
\textbf{(Proof)}
Since $\omega_{m} > 0$, 
$ \; \ms{D} (H_{\ph}(m)) \subset \ms{D} (N_{\ph})$ follows. Hence we see that $ \Psi_{m} \in \ms{D} (I \otimes N_{\ph})$. 
Let
\begin{equation}
T^{r,j}(f)=  - I \otimes a_{r} (\omega f ) 
- \kappa_{\I} \int_{\Rthree} \chi_{\I} (\mbf{x} ) 
( f, \frac{\chi_{\ph} e_{r,\mbf{x}}^{j} }{\sqrt{2(2\pi )^{3}  \omega } } ) 
(\fint \otimes I ) \dx 
  \label{AH-p479}.
\end{equation}
By the commutation relation of $H$ and $a_{r}(f) $, we see that
 for all $ \Phi \in \ms{D}_{0}$,
$ ( I \otimes a_{r} (f) \Psi_{m} , (H-E_{0} (H(m))) \Phi ) 
= (   T^{r,j} (f) \Psi_{m}  , \, \Phi )$. Hence, $ I \otimes a_{r} (f) \Psi_{E}  \in \ms{D} (H^{\ast})  =\ms{D} (H) $ and 
\begin{equation}
(H-E_{0}(H(m))) I \otimes a_{r} (f) \Psi_{m} = T^{r,j} (f) \Psi_{m}
\label{11/20.1}
\end{equation}
 follow. 
 Then by (\ref{11/20.1}),
\begin{align}
0 &\leq \; ( I \otimes a_{r}(f) \Psi_{m} ,(H (m) -E_{0}  
 (H(m))) (I \otimes a_{r} (f)) \Psi_{m} )   \notag \\
 &=-( I \otimes a_{r}(f) \Psi_{m}, \; 
I \otimes a_{r}(\omega_{m} f) \Psi_{m} ) \notag \\
 &\quad  -\kappa_{\I} \sum_{j} \int_{\Rthree} \chi_{\I} (\mbf{x} )
( f, \frac{\chi_{\ph}e_{r,\mbf{x}}^{j} }{\sqrt{2(2\pi )^{3}  \omega } } ) 
( I \otimes a_{r}(f) \Psi_{m},(\psi^{\ast} (\mbf{x} ) \alpha^{j}  \psi (\mbf{x} ) \otimes I ) \Psi_{m} ) \,  \dx .
 \label{AH-4.1}
\end{align}  
Let $ \{ g_{k} \}_{k=1}^{\infty}  $ be a complete orthonormal system of $\eltwo $ such that 
 $  g_{k} \in   \domain{\omega_{m}^{1/2} } \, , k \geqslant 1 $.   
By (\cite{AH97}; Lemma 4.2), it is seen that for all $ \Psi \in \ms{D} (I \otimes H_{\ph}(m) ) $,
\begin{equation}
\sum_{k=1}^{\infty} \sum_{r=1,2} 
( I \otimes a_{r} (\frac{g_{k}}{\sqrt{\omega_{m}}}) \Psi
, I \otimes a_{r}(\sqrt{\omega_{m}} g_{k}) \Psi ) =
 \| I \otimes N_{\ph}^{1/2} \Psi \|^{2}. \label{CONSNphhalf}
\end{equation}
By (\ref{AH-4.1}), it follows that for $\; N < \infty $,
\begin{align}
&\sum_{k=1}^{N}\sum_{r=1,2} \left( ( I\otimes a_{r}(\frac{g_{k}}{\sqrt{\omega_{m}}}) \Psi_{m},I\otimes a_{r}(\sqrt{\omega_{m}}g_{k}) \Psi_{m}) \right. \notag \\ 
& \left.  \quad  +\kappa_{\I}  \sum_{j=1}^{3}  \int_{\Rthree} \chi_{\I} (\mbf{x} )
( I \otimes a_{r}( \eta^{r,j,\mbf{x}}_{k} ) \Psi_{m},(\psi^{\ast} (\mbf{x} ) \alpha^{j}  \psi (\mbf{x} ) \otimes I ) \Psi_{m} ) \,  \dx   \right) \;  \leq 0\, ,
\label{AH-4.1.1}
\end{align}
where $
\eta^{r,j,\mbf{x} }_{k} = \frac{1}{\sqrt{\omega_{m}}} 
( g_{k} , \frac{\chi_{\ph} e_{r,\mbf{x}}^{j} }{\sqrt{2(2\pi )^{3}    \omega_{m}  \omega } } ) g_{k}$.
For $\;  N \in \mbf{N}$, we define  
\[
\lambda_{N}^{j} (\mbf{x} ) := \sum_{k=1}^{N}  \chi_{I} 
(\mbf{x} )
( I \otimes a_{r}( \eta^{r,j,\mbf{x}}_{k} ) \Psi_{m},(\fint \otimes I ) \Psi_{m} ) \, ,  \qquad j=1,2,3,
\]
and let
\[
\lambda^{j} (\mbf{x} ) : =\chi_{\I}(\mbf{x} ) \left( I \otimes a_{r}(\frac{\chi_{\ph}e_{r,\mbf{x}}^{j} }{\sqrt{2(2\pi )^{3}  \omega_{m}^{2} \omega } }  ) \Psi_{m},(\psi^{\ast} (\mbf{x} ) \alpha^{j}  \psi (\mbf{x} ) \otimes I ) \Psi_{m} \right) \,  , \quad j=1,2,3.
\]
Since $ \{ g_{k} \}_{k=1}^{\infty}  $ is a complete orthonormal system of $\eltwo $, we have 
\[
\lim_{N\to \infty} \left\| \sum_{k=1}^{N} \eta_{k}^{r,j,\mbf{x} } - \frac{\chi_{\ph}e_{r,\mbf{x}}^{j} }{\sqrt{2(2\pi )^{3}  \omega^{2}_{m} \omega } } \right\| =0 ,
\] 
for each $\mbf{x}$. 
Then
\begin{align*}
&|\lambda_{N}^{j} (\mbf{x})- \lambda^{j} (\mbf{x}) | \\
&\leq| \chi_{\I}(\mbf{x} ) ( I \otimes a_{r}(\sum_{k=1}^{N}
 \eta^{r,j,\mbf{x}}_{k} - \frac{\chi_{\ph}e_{r,\mbf{x}}^{j} }{\sqrt{2(2\pi )^{3}  \omega_{m}^{2} \omega } }  ) \Psi_{m},(\psi^{\ast} (\mbf{x} ) \alpha^{j}  \psi (\mbf{x} ) \otimes I ) \Psi_{m} )|\\
&\leq | \chi_{\I}(\mbf{x} )|
\| \sum_{k=1}^{N} \eta_{k}^{r,j,\mbf{x} } - 
\frac{\chi_{\ph}e_{r,\mbf{x}}^{j} }{\sqrt{2(2\pi )^{3}  \omega^{2}_{m} \omega } } \| \| I \tens N_{\ph}^{1/2} \Psi_{m} \| 
\| \psi^{\ast} (\mbf{x} ) \alpha^{j}  \psi (\mbf{x} ) \otimes I  \Psi_{m} \| \to 0 
\end{align*}
as $N \to \infty$.
We have also see  that
\[
\int_{\Rthree} |\lambda^{j}(\mbf{x})| \dx \leq
\| \chi_{\I} (\mbf{x}) \|_{L^{1}} \sum_{l,l'} 
| \alpha^{j}_{l,l'}| M_{l}^{\el}M_{l'}^{\el} \| \frac{\chi_{\ph}e_{r}^{j} }{\sqrt{2(2\pi )^{3}  \omega^{2}_{m}  \omega} } \| \| I \tens N_{\ph}^{1/2} \Psi_{m} \| 
\|  \Psi_{m} \| .
\]
Then  
$\; \lim_{N \to \infty} \int_{\Rthree} \lambda_{N}^{j} (\mbf{x} ) \dx =  \int_{\Rthree} \lambda_{N}^{j} (\mbf{x} )
\dx $ 
by the Lebesgue dominated convergence theorem.
 Therefore, by taking  $N$ to $\infty$  in (\ref{AH-4.1.1}), 
\begin{align*}
&\sum_{k=1}^{\infty} \sum_{r=1,2} 
( I \otimes a_{r} (\frac{g_{k}}{\sqrt{\omega_{m}}}) \Psi
, I \otimes a_{r}(\sqrt{\omega_{m}} g_{k}) \Psi ) \\
&\qquad  +\kappa_{\I} \int_{\Rthree}
\chi_{I}(\mbf{x} ) ( I \otimes a_{r}(\frac{\chi_{\ph}e_{r,\mbf{x}}^{j} }{\sqrt{2(2\pi )^{3}  \omega_{m}^{2} \omega } }  ) \Psi_{m},(\psi^{\ast} (\mbf{x} ) \alpha^{j}  \psi (\mbf{x} ) \otimes I ) \Psi_{m} ) \dx \leq 0 .
\end{align*}
By (\ref{CONSNphhalf}) , 
\begin{align*}
 \| (I \otimes N_{\ph}^{1/2} ) \Psi_{m} \|^{2}  
 &\leq | \kappa_{I} | \sum_{j=1}^{3} \left| 
 \int_{\Rthree} \chi_{\I}(\mbf{x} ) ( I \otimes a_{r}(\frac{\chi_{\ph}e_{r,\mbf{x}}^{j} }{\sqrt{2(2\pi )^{3} 
 \omega_{m}^{2} \omega } }  ) \Psi_{m},(\psi^{\ast} (\mbf{x} ) \alpha^{j}  \psi (\mbf{x} ) \otimes I ) \Psi_{m} ) \dx 
 \right| \\
&\leq  | \kappa_{\I}  | \| \chi_{\I} \|_{L^{1}} \sum_{j,l,l',r} | a_{l,l'}^{j} | M_{l}^{\el}M_{l'}^{\el} 
\left\|   \frac{\chi_{\ph}e^{j}_{r, \mbf{0}}}{\sqrt{2(2\pi )^{3} \omega_{m}^{2} \omega}} \right\|
\| (I \otimes N_{ph}^{1/2} ) \Psi_{m} \| \| \Psi \| .
\end{align*}
Thus, the proof is completed. $\blacksquare $

$\quad$ \\
Let  
\[
\ms{F}_{\el , \delta} :=E_{H_{\el}} ( [0,\delta) ) \ms{F}_{\el} , \quad \quad \delta >0 .
\]
We define the orthogonal projections by
\[
P_{\delta} = E_{E_{\el}}( [0,\delta )), \qquad
P_{\delta}^{\bot} = I - P_{\delta} ,
\qquad P_{\Omega_{\ph}} : \ms{F}_{\ph} \to \ms{L} \{   z \Omega_{\ph} \; | \; z \in \mbf{C}  \}.
\]
It can be proven, in similar manner to
Lemma \ref{Helboundness}, that
there exist constants $\; a_{2}>0 \; $ and $ \;  b_{2}>0 \; $ independent of $m$ such that
\begin{equation}
\| H_{0} (m) \Psi \| \leq a_{2} \| H(m) \Psi \| + b_{2} \| \Psi \| , 
\quad \quad \Psi \in \ms{D} (H(m)) . \label{Hmboundness}
\end{equation}

\begin{lemma}  \label{AH-p486}
Suppose  \textbf{(A.1)} -\textbf{(A.6)}. Let $\kappa_{\II}$ be sufficiently small such that $|\kappa_{\II} | \| H'_{\II} \| < \delta $. Then for $\epsilon >0$, 
\[
\| ( P_{\delta}^{\bot}  \otimes P_{\Omega_{\ph}}  ) \Psi_{m} \| 
\leq   \frac{ | \kappa_{\I} | \nu_{\epsilon} + |\kappa_{\II} | \| H'_{\II} \|}{\delta -E_{0} (H(m))}  \| \Psi_{m} \|  ,  
\]
where $ \nu_{\epsilon} = \epsilon L_{\I} (a_{2}  E_{0} (H(m)) + b_{2} ) + c_{\epsilon}L_{\I} + R_{\I} $ and 
$ c_{\epsilon}$ is the constant in (\ref{halfepsilon}). 
\end{lemma}
\begin{remark}
It is noted that $\; E_{0} (H(m) ) \; < \delta \;$ follows for 
sufficiently small $\kappa_{\II}$ as $ |\kappa_{\II} | \| H'_{\II} \| < \delta $ by Lemma \ref{12/15.5}.
\end{remark}
 \textbf{(Proof)}
 It is similar to (\cite{AH97} ; Lemma 4.7). 
 $\blacksquare$

$\quad $ \\ 
\begin{flushleft}
\textbf{( Proof of Theorem 1.2)}  
\end{flushleft}
By  the general theorem (\cite{AH97} ;  Lemma 4.9), it is enough to show that
$\lim_{m \to 0} E_{0} (H_{m}) =E_{0} (H)  \; $  
and there exists a nonzero weak limit of $\Psi_{m} $. 
We see that
$ E_{0} (H(m)) = ( \Psi_{m} , H(m) \Psi_{m}) 
=( \Psi_{m} , H \Psi_{m}) + m( \Psi_{m} , 
I \otimes N_{\ph}   \Psi_{m} ) $.
Then, we have $ \liminf_{m \to 0} E_{0} (H(m)) \geq E_{0}(H) $.
Since 
 $ \lim_{m \to \infty }  \| H(m) \Psi - H \Psi \| = 0\; $
  for $ \Psi \in \ms{D}_{0}$,  it is seen that 
  $H(m)$ converges to 
 $H$ as $m \to \infty$ in the strong resolvent sense. 
Hence, $ \limsup_{m \to 0}  E_{0} (H(m)) \leq E_{0} (H)  $.
Thus, we see that $\lim_{m \to 0}E_{0}(H(m)) = E_{0} (H) $.  
We next show that there exists a nonzero weak limit of $\Psi_{m} $. 
 We assume that $\| \Psi_{m} \| =1  $
for all $ m >0$. Then, there exists a subsequence
$ \{ \Psi_{m_{j}} \} $ such that $ \Psi := w-\lim_{j \to \infty} \Psi_{m_{j}}  \; $  exists.
  It is seen that $
P_{\delta} \otimes P_{\Omega_{\ph}}  = 
\geq I - I\otimes N_{\ph} - P_{\delta}^{\bot} \otimes P_{\Omega_{\ph}} $ follows. 
 By this inequality, Lemma \ref{AH-p482} and  
Lemma \ref{AH-p486} yield that  
\[
( \Psi_{m_{j}}, (P_{\delta} \otimes P_{\Omega_{\ph}})\Psi_{m_{j}})   
 \geq   1 
- \left( |\kappa_{\I} | \sum_{j,l,l',r} \nu^{j,l,l'}  \|   \frac{\chi_{\ph}e^{j}_{r}}{\sqrt{2(2\pi )^{3} \omega_{m_{j}}^{2} \omega }} \| \right)^{2}
 -  \left(\frac{ | \kappa_{\I} |  \nu_{\epsilon}  
+|  \kappa_{\II} | \| H_{\II}' \|  }{\delta -E_{0} (H(m_{j}))} \right)^{2}  . 
\]
Assume that 
$ | \kappa_{\II} | \| H'_{\II} \| \; < \; \delta \;  <\;  M $.  
 Then, $P_{\delta} \otimes P_{\Omega_{ph}} $ is a finite rank operator,  since $ \; \sigma (H_{el} ) = \{ 0 \} \cup
 [M ,\infty )  $. 
Taking the limit of  $\Psi_{m_{j}}$ as $\; j \to \infty $ in the above inequality, it follows that  
 $\; (\Psi_{0}, (P_{\delta} \otimes P_{\Omega_{\ph}})\Psi_{0} ) \; >0 \;$
 for sufficiently small
 $\kappa_{\I}$ and $\kappa_{\II}$. Then $ \Psi_{0} \ne 0$, and the proof is completed. $\blacksquare$

\subsection{Uniqueness of Ground States}
\begin{lemma} \label{statiophaseboson}
Assume \textbf{(A.7)}. Then, for $ \xi \in C^{\infty} (\Rthree )$, there exist $C_{r,j}^{1} > 0 \; $ and 
$ C_{r,j}^{1} \; > \,0  \; $ such that
\[
\left| \frac{}{} \right.
\left( \xi ,  \frac{\chi_{\ph}e^{j}_{r, \mbf{x}} 
e^{-it \omega} }{\sqrt{2(2\pi )^{3} \omega }} \right)
\left. \frac{}{} \right| \leq 
\frac{c_{r,j}^{1}}{t(1+t)} + |\mbf{x} | 
\frac{c_{r,j}^{2}}{t(1+t)} .
\]
\end{lemma}
\textbf{(Proof)}
It is seen that $ 
e^{- it \omega (\mbf{k} ) } = 
\frac{1}{( -it )} \frac{\omega (\mbf{k} )}{ k_{\nu }} \frac{\partial}{\partial k_{\nu}}
e^{-it \omega (\mbf{k}) } $. 
Using integration by parts, we obtain 
\begin{align*}
&\int_{\Rthree} \overline{\xi (\mbf{k} )} 
 \frac{\chi_{\ph} (\mbf{k})e^{j}_{r}(\mbf{k})}{\sqrt{2(2\pi )^{3} \omega ( \mbf{k}) }}
e^{i ( \mbf{k} \cdot \mbf{x} -  t \omega (\mbf{k} ))} \dk \\
&= \frac{1}{it} \int_{\Rthree}
\left( \frac{\partial}{\partial k_{\nu}} K_{r,j} (\mbf{k} ) \right) 
e^{i ( \mbf{k} \cdot \mbf{x} -  t \omega (\mbf{k} ) ) }\dk 
 -  \frac{x_{\nu} }{t} \int_{\Rthree} K_{r,j} (\mbf{k} ) 
e^{i( \mbf{k} \cdot \mbf{x} -  t \omega (\mbf{k} ) ) }\dk ,
\end{align*}
where $ 
 K_{r,j} (\mbf{k} ) =  \overline{\xi(\mbf{k} )}  
\frac{ \sqrt{\omega(\mbf{k})} \chi_{\ph} (\mbf{k} ) e^{j}_{r}(\mbf{k} ) }{\sqrt{2(2\pi )^{3} }  k_{\nu} } $.
Since $ e_{r}^{j} \in C^{\infty} (\Rthree \backslash O_{\ph}) $ and $ \chi_{\ph} \in C^{\infty} (\Rthree ) $,
 it follows that $ \chi_{\ph}  e_{r}^{j} \in C^{\infty} (\Rthree \backslash O_{\ph})  $. Hence for   
$ \, \xi \in  C^{\infty} (\Rthree \backslash \{ \mathbf{0} \} )$,  $ \; K_{r,j} , \; \partial_{k_{\nu}} K_{r,j} \; \in  C^{\infty }_{0} (\Rthree \backslash 
  O_{\ph} )   $. 
By (\cite{RS};Theorem XI.19),  there exist $c_{r,j}^{1}>0$
 and $c_{r,j}^{2} \; > \,0$ such that
\[
\sup_{\mbf{x} \in \Rthree } \left| 
\int_{\Rthree} ( \partial_{k_{\mu} }  K_{r,j} (\mbf{k} ) ) 
e^{i (\mbf{k} \cdot \mbf{x} -  t \omega (\mbf{k} )) }
d \mbf{k} \right| \leq \frac{c^{1}_{r,j} }{ 1+t } , 
\quad \sup_{\mbf{x} \in \Rthree } \left| 
\int_{\Rthree}  K_{r,j} (\mbf{k} )  
e^{i (\mbf{k} \cdot \mbf{x} -  t \omega (\mbf{k} )) }
d \mbf{k} \right| \leq \frac{c^{2}_{r,j} }{ 1+t } ,
\]
and hence the proof is completed. $\blacksquare$

$\quad$ \\ 
\textbf{(Proof of Proposition \ref{UNIQUENESSOFGS})} \\ 
Let  $\Phi, \; \Psi \; \in \ms{D} (H) $ and 
 $\xi \in C^{\infty} (\Rthree ) $. 
 Let us define the bilinear form
 $  [ X,Y ]^{0} : \ms{X} \times \ms{X} \; \to \; \mbf{C}$
\;  by
$ [ X,Y ]^{0} ( \phi , \psi ) =  (X^{\ast} \phi, Y \psi ) - 
 ( Y^{\ast} \phi ,  X \psi )$.
Then, we see that 
\begin{align*}
 [ H'( \kappa_{\I}  , \kappa_{\II} )   ,
 \;I \otimes a_{r} (\xi )  ]^{0} (  \Phi , \Psi  )  
&= \kappa_{\I} \sum_{j} \int_{\Rthree} \chi_{\I} (\mbf{x} ) ( \xi , 
\frac{\chi_{\ph} e^{j}_{r, \, \mbf{x}}}{\sqrt{2(2\pi )^{3} 
\omega}} )
( \Phi,   (\fint  \otimes I ) \Psi )  \dx   \\
&=\int_{\Rthree} \overline{\xi (\mbf{k} )} (  \Phi , T (r, \mbf{k} )\Psi  ) \dk ,
\end{align*}
where
\[
T(r, \mbf{k}) \Psi = \sum_{j}
\int_{\Rthree} \chi_{\I}(\mbf{x}) \frac{\chi_{\ph} (\mbf{k})e^{j}_{r}(\mbf{k})e^{i \mbf{k}\cdot \mbf{x} } }{\sqrt{2(2\pi )^{3} \omega (\mbf{k})}}
(\fint \tens I) \Psi \dx .
\]
It is sufficient to show that $H=H_{0} + H'( \kappa_{\I}  , \kappa_{\II} ) $ and $T(r,\mbf{k} )$ satisfy the
assumptions \textbf{(H.1)}-\textbf{(H.6)} in the appendix
 with $ X= X_{0} + q X' $ and $ S(r , \mbf{k}) $
 replaced by $ H=H_{0} + H'( \kappa_{\I}  , \kappa_{\II} ) $
 and $ T(r, \mbf{k} ) $, respectively.
 We  immediately see that $H=H_{0} + H'( \kappa_{\I}  , \kappa_{\II} )  $ and $T(r,\mbf{k} )$
satisfy \textbf{(H.1)}- \textbf{(H.3)} and \textbf{(H.5)}. 
Then the remaining task is to show
 \textbf{(H.4)} and \textbf{(H.6)}.
Let $\Psi $, $\Phi \; \in \ms{D}(H)$.
Then we see that 
\begin{align*}
&\int_{\Rthree} \overline{\xi (\mbf{k} )} (  \Phi ,  
e^{-it (H-E_{0}(H) + \omega (\mbf{k}) )} 
T (r, \mbf{k} )\Psi  ) \dk \\
&=\kappa_{\I} \sum_{j} \int_{\Rthree} \chi_{\I} (\mbf{x} ) ( \xi , 
\frac{\chi_{\ph} e^{j}_{r, \, \mbf{x}}e^{-it\omega } }{\sqrt{2(2\pi )^{3} 
\omega}} )
( e^{it(H-E_{0}(H))} \Phi,   (\fint  \otimes I ) \Psi ) \dx .
\end{align*}
Then by Lemma \ref{statiophaseboson}, and  \textbf{(A.3)}, \textbf{(A.6)},  we have
\begin{align*}
&\left| \int_{\Rthree} \overline{\xi (\mbf{k} )} (  \Phi , e^{-it(H -E_{0} (H) + \omega (\mbf{k} ) )}
T (r, \mbf{k} ),  \Psi  ) \dk \right| \\
&\leq 
\sum_{j,l,l'} |  \alpha^{j}_{l,l'} | M_{l}^{\el} M_{l'}^{\el}
 \| \Phi \| \; \| \Psi \| \; 
 \left( c_{r,j}^{1}  \| \chi_{\I} \|_{L^{1}}   + 
c^{2}_{r,j} \int_{\Rthree} | \mbf{x} |\; 
|  \chi_{\I} (\mbf{x} ) | \dx  \right)  
 \frac{1 }{ t( 1+t) } .
 \end{align*} 
Hence, we obtain $ \; \int_{\Rthree} \overline{\xi (\mbf{k} )} (  \Phi , e^{-it(H -E_{0} (H) + \omega (\mbf{k} ) )}
T (r, \mbf{k} ),  \Psi  ) \dk  \in L^{1}( [ 0, \infty ), dt)$.
It is also seen that
\[
 \| T(r, \mbf{k} ) \Psi \|  \leq 
\| \chi_{\I} \|_{L^{1}}  \left(  \sum_{j,l,l'} |  \alpha^{j}_{l,l'} | M_{l}^{\el} M_{l'}^{\el}  
\frac{ | \chi_{\ph} (\mbf{k})e^{j}_{r}(\mbf{k}) |}{\sqrt{2(2\pi )^{3} \omega (\mbf{k})}}   \right) \| \Psi  \|,
\]
and hence, $ \int_{\Rthree} \| T(r,\mbf{k} ) \Psi \|^{2} \dk < \infty $. Therefore, \textbf{(H.4)} follows.
It is  seen that for any ground state $\Psi_{g}$ of $H$,
\[
 \|  (H-E_{0} (H) + \omega (\mbf{k} ) )^{-1} T(r, \mbf{k} )
  \Psi_{g} \| 
 \leq   \left(  \| \chi_{\I} \|_{L^{1}}  \sum_{j,l,l'} |  \alpha^{j}_{l,l'} | M_{l}^{\el} M_{l'}^{\el}  
\frac{ | \chi_{\ph} (\mbf{k})e^{j}_{r}(\mbf{k}) |}{\sqrt{2(2\pi )^{3} \omega (\mbf{k})^{3} }}   \right) \| \Psi_{g}  \| .
\]
 Hence \textbf{(H.6)} follows.  $\blacksquare $

$\quad$ \\

\section{Asymptotic fields}
\subsection{Existence of asymptotic fields}
Let
\[
a_{r,t}^{\sharp}(\xi) = e^{it H}
e^{-itH_{0}}( I \otimes a_{r}^{\sharp}(\xi))e^{itH_{0}}
e^{-it H} ,
\]
and 
\[
b_{s,t}^{\sharp}(\eta) = e^{it H}
e^{-itH_{0}}( b_{s}^{\sharp}(\eta ) \tens I )e^{itH_{0}}
e^{-it H} ,   \qquad 
 d_{s,t}^{\sharp}(\zeta) = e^{it H}
e^{-itH_{0}}( d_{s}^{\sharp}(\zeta ) \tens I )e^{itH_{0}}
e^{-it H} ,
\]
where $ X^{\sharp} = X$ or $ X^{\ast}$. \\

$\quad$ \\ 
It is proven, in a manner similar to that used in the proof of Lemma \ref{Helboundness}, that there exist $a_{0}>0$, $b_{0} \; >0$ such that
\begin{equation}
\| H_{0} \Psi \| 
\leq a_{0} \| H \Psi \|  + b_{0} \| \Psi \| .
\label{HzeroHbound}
\end{equation}
\textbf{(Proof of Theorem \ref{asympphoton})}  \\ 
Let $\xi \in C_{0}^{\infty} (\Rthree \backslash O_{\ph} ) $. 
We see that $ e^{-itH_{0}} I \tens a_{r}(\xi ) e^{itH_{0}}
=I \tens a_{r}( e^{-it\omega} \xi ) $.
Let $ \; \Phi(t)= e^{-itH}\Phi \; $ and $ \; \Psi(t)= e^{-itH}\Psi \; $
 for $\Phi, \, \Psi \in \ms{D}(H) $.
By the strong differentiability of $e^{itH} \Psi \; $ and $e^{itH_{0}} \Psi $   with respect to $t$, 
\begin{align*}
&( \Phi,a_{r,T}(\xi )\Psi )  - ( \Phi,a_{r,T_{0}}(\xi )\Psi ) \\
 &\quad = \int_{T_{0}}^{T} \left\{
-\kappa_{\I} \sum_{j}
 \int_{\Rthree} \chi_{\I}(\mbf{x})( e^{-it\omega} \xi ,\;
f_{r,\mbf{x}}^{j}    )_{L^{2}} \, ( \Phi(t), (\fint \tens I) 
\Psi (t) )  \dx  \right\} d t .
\end{align*}
By Lemma \ref{statiophaseboson} and \textbf{(A.6)}, 
\begin{align}
 & \left| \kappa_{\I} \sum_{j}
 \int_{\Rthree} \chi_{\I}(\mbf{x})( e^{-it\omega} \xi ,\;
f_{r,\mbf{x}}^{j}    )_{L^{2}} \, ( \Phi(t), (\fint \tens I) 
\Psi (t) )  \dx \right|   \notag  \\
&\leq   |\kappa_{\I}| \| \Phi\| \| \Psi \|  \sum_{j,l,l'}
|\alpha_{l,l'}^{j}|M_{l}^{\el}M_{l'}^{\el}
\left( c_{r,j}^{1}  \| \chi_{\I} \|_{L^{1}} + c_{r,j}^{2} \int_{\Rthree} | \mbf{x} | |\chi_{\I} (\mbf{x}) | \dx  
 \right) \frac{1}{t(1+t)} .
\end{align}
Hence $
\| a_{r,T} (\xi ) \Psi - a_{r,T'} (\xi ) \Psi \| \leq
 \text{const.} \int_{T'}^{T}\frac{1}{t(1+t)} dt \to 0 $, 
as $T, T'\, \to \, \infty$.
Thus, for $\; \xi \in C_{0}^{\infty} (\Rthree \backslash O_{\ph}
 ) $, 
$  a_{r, \infty} (\xi ) \Psi = s-\lim_{t \to \infty} a_{r,t} (\xi ) \Psi   \;$  exists for $\Psi \in \domain{H} $.
Let  $ f \in  \domain{\omega^{-1/2}} $. 
Since $C_{0}^{\infty}( \Rthree \backslash O_{\ph} ) $  is a core of $\omega^{-1/2}$, there exists a sequence $\{ f_{n} \} \, $ $\subset$  $C_{0}^{\infty}( \Rthree \backslash O_{\ph}
 ) \,$ such that $ \| f_{n} - f \| \to 0 $,  
and $ \| \omega^{-1/2}f_{n} -  \, \omega^{-1/2}  f \| \to 0 $ as $n \to 0$.  Then for  $t' < t$, 
\begin{align}
\| a_{r,t} (f) \Psi - a_{r,t'} (f) \Psi \|   
 \leq   
& \| a_{r} (e^{-it\omega}  ( f-f_{n} )) e^{-itH} \Psi \|     
+ \| a_{r,t} (f_{n}) \Psi - a_{r,t'} (f_{n}) \Psi \|   \notag \\
& \quad +  \| a_{r} (e^{-it'\omega}  ( f-f_{n} )) e^{-it'H} \Psi \| .
 \label{8/21.1}
\end{align}
By (\ref{boundar}), (\ref{halfepsilon}), and 
(\ref{HzeroHbound}),
\begin{equation}
 \| a_{r} (e^{-it\omega}  ( f-f_{n} )) e^{-itH} \Psi \| 
\leq \| \frac{f-f_{n}}{\sqrt{\omega}} \|
( \epsilon a_{0} \| H \Psi \| +  (\epsilon b_{0}+ c_{\epsilon} )  \| \Psi \| ) +  \| f-f_{n} \| \| \Psi \|  \to 0  ,
\end{equation}
as $n \to \infty$.
Hence by (\ref{8/21.1}),
$ \| a_{r,t} (f) \Psi  - a_{r,t'} (f) \Psi   \| \to 0 $,
 as $t, t' \to \infty$.  $\blacksquare$

$\quad $ \\
\begin{lemma} \label{originalone} 
Let $\eta, \; \zeta \in \eltwo$. Then
\begin{align}
&[ \psi_{l}^{\ast}(\mbf{x}) \, \psi_{l'}(\mbf{x})  , \; b_{\tau }(\eta)  ] = -( \eta,g_{\tau, \mbf{x}}^{l}  )
 \psi_{l'}(\mbf{x})   , \\
&[ \psi_{l}^{\ast}(\mbf{x}) \, \psi_{l'}(\mbf{x})  , \; d_{\tau}(\zeta)  ] = ( \zeta ,h_{\tau , \mbf{x}}^{l'}  )  \psi_{l}^{\ast}(\mbf{x}) .
\end{align}
\end{lemma}

$\quad$ \\ 
By Lemma \ref{originalone}, it follows that 
\begin{align}
&[ \fint , b_{s}(\eta ) ] = - \sum_{l,l'} \alpha_{l,l'}^{j}
( \eta, g_{s,\mbf{x}}^{l}  ) \;  \psi_{l'}(\mbf{x}) ,  \label{ori1}
\\
&[ \fint , d_{s}(\zeta ) ] = \sum_{l,l'} \alpha_{l,l'}^{j}
( \zeta,h_{s ,\mbf{x}}^{l'}  ) \; \psi_{l}^{\ast}(\mbf{x}) , \label{ori2}
 \\
&[ \rho (\mbf{x}) \rho (\mbf{y}), \; b_{s}(\eta ) ] 
= -\sum_{l} \left( ( \eta , g_{s, \mbf{y}}^{l}  )
\rho (\mbf{x}) \psi_{l}(\mbf{y}) 
+ ( \eta , g_{s, \mbf{x}}^{l}  ) \; \psi_{l}(\mbf{x})
 \rho (\mbf{y}) \right) ,   \label{ori3} \\
&[ \rho (\mbf{x}) \rho (\mbf{y}), \; d_{s}(\eta ) ] 
= \sum_{l} \left( ( \zeta , h_{s, \mbf{y}}^{l}  )
\rho (\mbf{x}) \psi_{l}^{\ast}(\mbf{y}) 
+ ( \zeta , h_{s, \mbf{x}}^{l}  ) \;  \psi_{l}^{\ast}(\mbf{x})
 \rho (\mbf{y}) \right) . \label{ori4}
\end{align}
 
$\quad$ \\  
\textbf{(Proof of Lemma \ref{originalone})} \\ 
We see that by the anti-commutation relations 
\begin{align*}
&[ b_{s}^{\ast} (g_{s,\mbf{x}}^{l} ) b_{s'} (g_{s',\mbf{x}}^{l'} ), \; b_{\tau}(\eta ) ] 
= -\{ b_{s}^{\ast} (g_{s, \mbf{x}}^{l} ),  
b_{\tau}(\eta ) \} b_{s'} (g_{s', \mbf{x}}^{l'} ) 
=-(\eta, g_{s, \mbf{x}}^{l}  )
 \delta_{s,\tau} \; b_{s'} (g_{s', \mbf{x}}^{l'} ) ,\\
&[ b_{s}^{\ast} (g_{s, \mbf{x}}^{l} ) d^{\ast}_{s'} (h_{s',\mbf{x}}^{l'} ), \; b_{\tau}(\eta ) ] 
= -\{ b_{s}^{\ast} (g_{s, \mbf{x}}^{l} ), 
b_{\tau}(\eta ) \} d^{\ast}_{s'} (h_{s', \mbf{x}}^{l'} ) 
=-(\eta, g_{s, \mbf{x}}^{l}  ) \delta_{s,\tau} \; 
d^{\ast}_{s'} (h_{s', \mbf{x}}^{l'} ),  \\
& [ d_{s} (h_{\mbf{x}}^{l} ) b_{s'} (g_{s', \mbf{x}}^{l'} ), \; b_{\tau}(\eta ) ] = -\{ d_{s} (h_{s, \mbf{x}}^{l} ), 
b_{\tau}(\eta ) \} b_{s'} (g_{s', \mbf{x}}^{l'} ) =0 , \\
& [ d_{s} (h_{\mbf{x}}^{l} ) d^{\ast}_{s'} (h_{s', \mbf{x}}^{l'} ), 
\; b_{\tau}(\eta ) ] = -\{ d_{s} (h_{s, \mbf{x}}^{l} ), 
b_{\tau}(\eta ) \} d^{\ast}_{s'} (h_{s', \mbf{x}}^{l'} ) =0 .
\end{align*}
Hence, 
\[
[ \psi_{l}^{\ast}(\mbf{x}) \, \psi_{l'}(\mbf{x})  , \; b_{s}(\eta)  ] 
=- (\eta, g_{\tau , \mbf{x}}^{l}  ) \sum_{s'}
( b_{s'} (g_{s', \mbf{x}}^{l'} ) + d^{\ast}_{s'} (h_{s',\mbf{x}}^{l'} )) \\
=- (\eta , g_{\mbf{x}}^{l}  ) \psi_{l'}(\mbf{x}) .
\]
Similarly, we can obtain 
\begin{align*}
&[ b_{s}^{\ast} (g_{s,\mbf{x}}^{l} ) d^{\ast}_{s'} (h_{s', \mbf{x}}^{l'} ), \; d_{\tau}(\zeta ) ] = (\zeta, h_{s', \mbf{x}}^{l'}  ) 
\delta_{s' ,\tau} b_{s}^{\ast} (g_{s, \mbf{x}}^{l} )  , \\ 
  &[ d_{s} (h_{s, \mbf{x}}^{l} ) d^{\ast}_{s'} (h_{s', \mbf{x}}^{l'} ), \; d_{\tau}(\zeta ) ] = (\zeta, h_{s', \mbf{x}}^{l'}  ) 
\delta_{s' ,\tau} d_{s} (h_{s, \mbf{x}}^{l} ) ,  \\
& [ b_{s}^{\ast} (g_{s, \mbf{x}}^{l} ) b_{s'} (g_{s', \mbf{x}}^{l'} ), \; d_{\tau}(\zeta ) ] =
 [ d_{s} (h_{s, \mbf{x}}^{l} ) b_{s'} (g_{s', \mbf{x}}^{l'} ), \; d_{\tau}(\zeta ) ] =0 . 
\end{align*}
Hence,  $ [ \psi_{l}^{\ast}(\mbf{x}) \, \psi_{l'}(\mbf{x})  , \; d_{s}(\zeta)  ]=  (\zeta, h_{ \tau , \mbf{x}}^{l'}  ) \psi_{l}^{\ast}(\mbf{x}) $. $\blacksquare$

$\quad $ \\ 
It is known by (\cite{RS}; Theorem XI.15) that
 for $ \eta \in C_{0}^{\infty} (\Rthree)$,
there exist constants $\nu_{l} (s)>0 $ and $\tilde{\nu}_{l}(s)>0 $
 such that
\begin{equation}
\sup_{\mbf{x} \in \Rthree} 
|(e^{itE_{M}}  \eta , g_{s,\mbf{x}}^{l} )_{L^{2}}| 
 \leq
\frac{\nu_{l} (s)}{(1+t)^{3/2}} , \quad
\sup_{\mbf{x} \in \Rthree} 
|(e^{itE_{M}}  \eta , h_{s, \mbf{x}}^{l} )_{L^{2}}| 
\leq \frac{\tilde{\nu}_{l}(s)}{(1+t)^{3/2}} .
\label{statiophasefermi}
\end{equation}
We also see from (\ref{halfepsilon}) and (\ref{boundAj}), 
 that for
 $\epsilon> 0$,
\begin{equation}
\| I \otimes A_{j} (\mbf{x}) \Psi  \| \leq
L_{I}^{j}(\epsilon ) \| H \Psi \| + 
R_{I}^{j}(\epsilon ) \|  \Psi \| ,
\label{AjHbound}
\end{equation} 
where $
L_{\I}^{j}(\epsilon ) = 2 \epsilon a_{0} 
\sum_{r} M_{2,j,r}^{\ph}  ,  \; 
R_{\I}^{j}(\epsilon )= \sum_{r}( 2 M_{2,j,r}^{\ph}  (\epsilon b_{0} +c_{\epsilon}) + M_{1,j,r}^{\ph} ) $.

$\quad$ \\ 
\textbf{(Proof of Theorem \ref{asympDirac})} \\
 Let $ \eta \in C_{0}^{\infty} (\Rthree \backslash 
O_{\el}  )$.
It is seen that $ e^{-itH_{0}}( b_{s}(\eta ) \tens I )e^{itH_{0}}
= b_{s} ( e^{-itE_{M}} \eta ) \tens I $. 
Let  $ \Phi(t)= e^{-itH}\Phi$ and $ \Psi(t)= e^{-itH}\Psi$,
  for  $\Phi , \, \Psi \in \ms{D}(H) $, respectively.
  As in the case of the photon fields, 
\begin{align}
& ( \Phi, b_{s,T}(\eta )\Psi ) -  ( \Phi, b_{s,T_{0}}(\eta )\Psi )
 \notag \\
&=
\int_{T_{0}}^{T} \left\{  
 \kappa_{\I} \;  [ H_{\I}'   ,  \; b_{s}( e^{-itE_{M}} \eta ) \tens I ]^{0} \; (  \Phi (t) , \Psi (t) )  
+  \kappa_{\II} \;  [ H_{\II}'   ,  \; b_{s}( e^{-itE_{M}} \eta ) \tens I ]^{0} \; (  \Phi (t) , \Psi (t) ) \right\}  dt .   
\end{align}
By (\ref{ori1}), 
\begin{equation}
 [ H_{\I}'   ,  \; b_{s}( e^{-itE_{M}} \eta ) \tens I ]^{0} \; (  \Phi (t) , \Psi (t) )  
= - \sum_{j,l,l'} \alpha_{l,l'}^{j} \int_{\Rthree} ( e^{-itE_{M}} \eta,
  g_{s, \mbf{x}}^{l}  )_{L^{2}} 
( \Phi (t) ,  \psi_{l'}(\mbf{x}) \tens A_{j} (\mbf{x}) 
\Psi (t) ) \dx  .
\end{equation}
We  also see that by (\ref{AjHbound}),
\[
| ( \Phi (t) ,  \psi_{l'}(\mbf{x}) \tens A_{j} (\mbf{x}) 
\Psi (t) ) | 
 \leq M_{l'}^{\el} \| \Phi \| 
( L_{\I}^{j}(\epsilon )   \| H \Psi \| + R_{\I}^{j}(\epsilon ) \| \Psi \|   ) ,   
\]
 and hence from (\ref{statiophasefermi})
\begin{equation}
\left| [ H_{\I}'   ,  \; b_{s}( e^{-itE_{M}} \eta ) \tens I ]^{0} \; (  \Phi (t) , \Psi (t) )  \right|
 \leq |\kappa_{\I}| \| \Phi \| 
\sum_{i,l,l'} |\alpha_{l,l'}^{j}| \nu_{l} M_{l'}^{\el}
(L_{\I}^{j} (\epsilon )  \| H\Psi \| + 
R_{\I}^{j}(\epsilon ) \| \Psi \| ) \frac{1}{(1+t)^{3/2}} .
 \label{11/20.5}
\end{equation}
In addition, we see that by (\ref{ori3}),
\begin{align*}
[ H_{\II}'   ,  \; b_{s}( e^{-itE_{M}} \eta ) \tens I ]^{0} \; (  \Phi (t) , \Psi (t) )  
&=-i  \sum_{l} \int_{\Rthree} \frac{\chi_{\II}(\mbf{x}) \chi_{\II}(\mbf{y}) }{|\mbf{x} -\mbf{y}|} 
\left\{  ( e^{-itE_{M}} \eta, g_{s, \mbf{y}}^{l}  )
( \Phi (t) , \rho (\mbf{x}) \psi_{l}(\mbf{y}) \tens I  \Psi (t) ) \right.  \\
&\qquad  \quad  \quad \left. + ( e^{-itE_{M}} \eta ,
  g_{s, \mbf{x}}^{l}  ) 
( \Phi (t) , \psi_{l}(\mbf{x})  \rho (\mbf{y}) \tens I \Psi (t) ) \right\} \dx \dy  .
\end{align*}
By (\ref{AjHbound}), 
\begin{equation}
 |( \Phi (t) ,  \rho (\mbf{x}) \psi_{l}(\mbf{y}) \tens I \Psi (t) ) |  
\leq M_{l}^{\el}  \sum_{\nu} (M_{\nu}^{\el})^{2}
  \| \Phi \|  \|  \Psi  \| 
\end{equation}
follows. Then by (\ref{statiophasefermi}), we have
\begin{equation}
\left| [ H_{\II}'   ,  \; b_{s}( e^{-itE_{M}} \eta ) \tens I ]^{0} \; (  \Phi (t) , \Psi (t) ) \right| 
\leq  2 M_{\II} \left( 
\sum_{l, \nu}   M_{l}^{\el}   (M_{\nu}^{\el})^{2}
  \| \Phi \|  \|  \Psi  \|  \right) 
\frac{1}{(1+t)^{3/2}}   . \label{11/20.6}
\end{equation}
 Thus, (\ref{11/20.5}) and  (\ref{11/20.6}) yield 
\[
\| b_{s,T} (\eta ) \Psi - b_{s,T'} (\eta ) \Psi \| \leq
 \text{const.} \int_{T'}^{T}\frac{1}{(1+t)^{3/2}} dt \to 0 ,
\]
as $T, T'\, \to \, \infty$.
Hence, we obtain the asymptotic fields 
$\, b_{s,\infty} (\eta )  := s-\lim_{t \to \infty}   b_{s,t} (\eta ) \Psi $  for $ \, \eta \in C_{0}^{\infty} (\Rthree ) $. Since $ C_{0}^{\infty} (\Rthree )$ is dense in $\eltwo$ and
$\| b_{s,t}(\eta) \| \leq \| \eta \|$, we can extend the asymptotic fields $\, b_{s,\infty} (\eta ) $ for $ \, \eta \in \eltwo $. The proof is thus completed.
 $\blacksquare$

\subsection{Basic Properties of the Asymptotic Fields}

\begin{lemma} \label{adjointasymp}
Assume  \textbf{(A.1)}-\textbf{(A.3)}, \textbf{(A.5)}, 
\textbf{(A.7)} and \textbf{(A.8)}. \\ 
\textbf{(1)}
Let $\,  \eta , \zeta  \in \eltwo $. Then for 
$\Phi, \; \Psi \; \in \domain{H} $,
\[ (\Phi, \; b_{s, \pm \infty} (\eta ) \Psi  )
= (   b_{s, \pm \infty}^{\ast} (\eta ) \Phi , \Psi ) , \qquad
(\Phi, \; d_{s, \pm \infty} (\zeta ) \Psi  )
= (   d_{s, \pm \infty}^{\ast} (\zeta ) \Phi , \Psi ) .
\]
\textbf{(2)} Let $\xi \in \domain{\omega^{-1/2}}$. 
Then for $\Phi, \; \Psi \; \in \domain{H} $,
\[
 (\Phi,\; a_{r, \pm \infty} (\xi ) \Psi  )
= (   a_{r, \pm \infty}^{\ast} (\xi ) \Phi , \Psi ).
\]
\end{lemma}
\textbf{(Proof)} \\
It is seen that 
\[
(\Phi, b_{s, \pm \infty} (\eta ) \Psi  )=
\lim_{t \to \pm \infty} (\Phi, b_{s, t} (\eta ) \Psi  )
= \lim_{t \to \pm \infty} (   b_{s,t}^{\ast} (\eta ) \Phi , \Psi ) = (   b_{s, \pm \infty}^{\ast} (\xi ) \Phi , \Psi ).
\]
Hence we have (1). Similarly, we can prove (2).
$\blacksquare$

$\quad$ \\ 
 \begin{lemma}  \label{asympanticommu}
Assume that \textbf{(A.1)}-\textbf{(A.3)}, \textbf{(A.6)},
\textbf{(A.7)}. Let
$\eta, \; \eta ' \, \zeta, \; \zeta '  \; \in \eltwo $.
 It follows that,
 \begin{align*}
&\{ b_{s,\pm \infty}(\eta ), 
b_{s', \pm \infty}^{\ast}(\eta' ) \}  
= \delta_{s,s'}(\eta , \eta') , \\ 
&\{ d_{s, \pm \infty}(\zeta ), d_{s,\pm \infty}^{\ast}(\zeta') \} = \delta_{s,s'}(\zeta , \zeta') , \\
&\{ b_{s, \pm \infty}(\eta ), b_{s',\pm \infty} (\eta ') \} = \{ d_{s,\pm \infty}(\zeta ), d_{s',\pm \infty} (\zeta ') \} =0 , \\  
& \{ b_{s,\pm \infty}(\eta), d_{s', \pm \infty} (\zeta ') \} = \{ b_{s, \pm \infty}(\eta ), d_{s', \pm \infty}^{\ast} (\zeta ') \} =0 .
\end{align*}
\end{lemma}
\textbf{(Proof)} 
 It is seen that for $\Psi , \Phi \in \ms{F}_{\QED}$
\begin{align*}
( \Phi , \{ b_{s, t }(\eta ) ,\; b_{s', t }^{\ast}(\eta ')   \} \Psi ) 
&=  ( e^{-itH }\Phi ,\; I \otimes \{  b_{s, t }
(e^{-it  E_{M} } \eta ) ,\; b_{s', t}^{\ast}(e^{-it E_{M}}
\eta ')   ] e^{-itH} \Psi ) \\
&= \delta_{s,s'} (\eta , \; \eta ') (\Phi, \Psi) .
\end{align*}
Hence we obtain $\{ b_{s,\pm \infty}(\eta ), 
b_{s', \pm \infty}^{\ast}(\eta' ) \}  
= \delta_{s,s'}(\eta , \eta')$. Similarly,  
it can be proven  in  other cases. $\blacksquare$

\begin{lemma}  \label{asympcommutation}
Assume  \textbf{(A.1)}-\textbf{(A.3)}, \textbf{(A.5)} and 
\textbf{(A.7)}.   Let
$\xi, \; \xi '   \; \in \domain{\omega^{k/2}}$, $k= -1,1,2$. 
 Then, on $ \domain{H}$,
\begin{align*}
& \textbf{(1)} \quad [a_{r, \pm \infty }(\xi ) ,\; 
a_{r', \pm \infty }^{\ast}(\xi ')   ]  = \delta_{r,r'}
(\xi , \; \xi ')  , \\
& \textbf{(2)} \quad [ \, a_{r, \pm \infty}(\xi ), \, a_{r', \pm \infty}(\xi ' )  ] = [ a_{r, \pm \infty}^{\ast}(\xi ), \, 
 a_{r', \pm \infty}^{\ast}(\xi ' ) ] =0 .
\end{align*}
\end{lemma}
\textbf{(Proof)}  
 It is similar to Lemma \ref{asympanticommu}. $\blacksquare$

\begin{lemma} \label{Weylrelation}
Assume  \textbf{(A.1)}-\textbf{(A.3)}, \textbf{(A.5)}, 
\textbf{(A.7)} and \textbf{(A.8)}. \\
\textbf{(1)} Let $\eta, \zeta \; \in \eltwo$.
 Then, for $\Psi \; \in \domain{H} $,
\[
e^{itH} b_{s, \pm \infty }^{\sharp} (\eta ) \Psi 
= b_{s, \pm \infty }^{\sharp} (e^{itE_{M}} \eta ) e^{itH} \Psi
 ,  \qquad
e^{itH} d_{s, \pm \infty }^{\sharp} (\zeta ) \Psi 
= d_{s, \pm \infty }^{\sharp} (e^{itE_{M}} \zeta ) e^{itH} \Psi .
\]
\textbf{(2)} Let $\xi \in \domain{\omega^{-1/2}}$. 
 Then, for $\Psi \; \in \domain{H} $,
\[
e^{itH} a_{r, \pm \infty }^{\sharp} (\xi ) \Psi 
= a_{r, \pm \infty }^{\sharp} (e^{it\omega} \xi ) 
e^{itH} \Psi  . 
\] 
\end{lemma}
\textbf{(proof)} \\ 
We see that
\[
e^{itH} b_{s,t'} (\xi ) \Psi
= e^{i(t+t')H } e^{-i(t+t')H_{0} }  \; 
 b_{s} (e^{it E_{M}} \eta ) \tens I \; e^{i(t+t')H_{0} } 
 e^{-i(t+t')H }  e^{itH}\Psi .
\]
By taking   $ t' \to   \pm \infty $, we obtain \textbf{(1)} . 
 We can prove \textbf{(2)} similarly to \textbf{(1)}.
$\blacksquare$

$\quad$\\ 
Since $ a^{\sharp}_{r} (\xi ) $ maps
$ \domain{H_{\ph}^{3/2}} $ to  
$ \domain{H_{\ph}}$, 
 it can be proven   
   that $ a^{\sharp}_{r, \pm \infty} (\xi ) $ maps
 $ \domain{|H|^{3/2}} $ to  $ \domain{H}$ in the similar way as
  (\cite{Hi01} ; Lemma 4.10, Lemma 4.11).
 Then by the strong differentiability of $e^{itH} \Psi$
 and Lemma \ref{Weylrelation}, we obtain the following lemma.
 
 \begin{lemma}  
 $\quad $ \\ 
Assume  \textbf{(A.1)}-\textbf{(A.3)}, \textbf{(A.5)}, 
\textbf{(A.7)} and \textbf{(A.8)}. \\ 
\textbf{(1)} Let $ \eta, \; \zeta \in \domain{E_{M}} $.
 It follows that on $ \domain{H}$,
 \begin{align*}
&[ H , \, b_{s, \pm \infty } (\eta ) ]
= -b_{s, \pm \infty } (E_{M} \eta ), \quad \quad 
 [ H , \, b^{\ast}_{s, \pm \infty } (\eta ) ]
= b^{\ast}_{s, \pm \infty } (E_{M} \eta ) ,  \\
&[ H , \, d_{s, \pm \infty } (\zeta ) ]
= -d_{s, \pm \infty } (E_{M} \zeta ), \quad \quad 
 [ H , \, d^{\ast}_{s, \pm \infty } (\zeta ) ]
= d^{\ast}_{s, \pm \infty } (E_{M} \zeta )  .
\end{align*}
\textbf{(2)}  Let $ \xi \in  \domain{\omega^{-1/2}}
 \cap \domain{\omega}  $. It follows that on 
 $\; \domain{|H|^{3/2}} $,
\[
[ H , \, a_{r, \pm \infty } (\xi ) ] 
= - a_{r, \pm \infty } (\omega \xi ) , \quad \quad
[ H , \, a^{\ast}_{r, \pm \infty } (\xi ) ] 
= a^{\ast}_{r, \pm \infty } (\omega \xi ) .
\]
\end{lemma}

\begin{lemma}  \label{newvacuum}
$\; $ \\ 
Assume  \textbf{(A.1)}-\textbf{(A.3)}, \textbf{(A.6)}-
\textbf{(A.8)}. Let $\Psi_{E}$ be an eigenvector of $H$ with the
eigenvalue $E$. Then   \\ 
\textbf{(1)}  for  $\eta, \; \zeta \in \eltwo $,
\[
b_{s, \pm \infty } (\eta ) \Psi_{E} = 0 , \qquad
d_{s, \pm \infty } (\zeta ) \Psi_{E} = 0 ,
 \]
 \textbf{(1)} for $ \xi \in  \domain{\omega^{-1/2}}  $,  
 \[
 a_{r, \pm \infty } (\xi ) \Psi_{E} = 0 .
 \]
\end{lemma}
\textbf{(Proof)}  \\ 
Let  $\eta \in C_{0}^{\infty} (\Rthree \backslash O_{\el} ) $. We see that 
 \[
 \| b_{s, \, t } (\eta ) \Psi_{E} \|
=  \| e^{itH}  b_{s} ( e^{-it E_{M}} \eta )  \tens I e^{-itH} 
 \Psi_{E} \| = \|  b_{s} ( e^{-it E_{M}}  \eta ) \tens I   
\Psi_{E} \|  .
 \]
 Let
$  \Phi =   b_{s_{1}}^{\ast} (\eta_{1} ) \cdots b_{s_{n}}^{\ast} (\eta_{n} )  d_{\tau_{1}}^{\ast} (\zeta_{1} ) \cdots d _{\tau_{n'}}^{\ast} (\zeta_{n'} ) \Omega_{\el} \tens  \Phi_{\ph}   
\; \in \ms{F}_{el}  ( C^{\infty}(\Rthree \backslash O_{\el}))
\hat{\otimes} \ms{F}_{\ph}^{\fin} ( \ms{D} (\omega ))  $. By the anti-canonical commutation relation,  
\begin{align*}
&\|  b_{s} ( e^{-it E_{M}} \eta )  \tens I
\Phi \|  \\
& \leq 
\sum_{j=1}^{n} | (e^{-itE_{M}} \eta , \eta_{j} )_{L^{2}} |
 \, \delta_{s, s_{j}} \,
 \| b_{s_{1}}^{\ast} (\eta_{1} ) \cdots 
 \widehat{b_{s_{j}}^{\ast} (\eta_{j} )} \cdots
b_{s_{n}}^{\ast} (\eta_{n} )  d_{\tau_{1}}^{\ast} (\zeta_{1} ) \cdots d _{\tau_{n'}}^{\ast} (\zeta_{n'} ) \Omega_{\el} \tens  \Phi_{\ph}
\| .
\end{align*}
By (\cite{RS} ; Theorem XI.19) there exists a constant $F_{j}$ such that
$ | (e^{-it E_{M}} \xi, \xi_{j} )_{L^{2}} | \leq 
\frac{F_{j}}{(1+t)^{3/2}} $. 
Hence, we have $\lim_{t \to \infty} \| I \tens a_{r} ( e^{-it\omega} \xi )  \Phi \|  =0 $. 
 Since $\ms{F}_{el}  ( C^{\infty}(\Rthree \backslash O_{\el}))
\hat{\otimes} \ms{F}_{\ph}^{\fin} ( \ms{D} (\omega ))$ 
 is a core of $H$, 
we obtain $\|   b_{s, \infty } (\eta ) \Psi_{E} \| =0 $
 for $\eta \in  C_{0}^{\infty} (\Rthree \backslash O_{\el} )$. 
 Since $C_{0}^{\infty} (\Rthree \backslash O_{\el}  ) $ is 
dense in $\eltwo $ and  $E_{M}^{-1/2}$ is a bounded operator,
 we can extend for all $ \eta \in \eltwo$. 
 Thus, we can complete the proof of \textbf{(1)}. 
  \textbf{(2)} is proven similarly to  \textbf{(1)}.
$\blacksquare$

$\quad$ \\
Let $\Psi_{g}$ be a ground state of $H$.
We next consider the asymptotic in$/$out-going Fock space.
Let
\begin{align*}
\ms{F}_{\pm \infty}^{n,l,m} =&
\ms{L}
\left\{
a_{r_{1}, \pm \infty}^{\ast} (\xi_{1})
\cdots a_{r_{n}, \pm \infty}^{\ast} (\xi_{n})
b_{s_{1}, \pm \infty}^{\ast} (\eta_{1})
\cdots b_{s_{l}, \pm \infty}^{\ast} (\eta_{l})
d_{\tau_{1}, \pm \infty}^{\ast} (\zeta_{1})
\cdots d_{\tau_{m}, \pm \infty}^{\ast} (\zeta_{m})
\Psi_{g} , \;  \right. \\  
 &| \left. \;  
 \xi_{i} \in \domain{\omega^{-1/2}}, \; 
 \eta_{j} , \; \zeta_{k} \in \eltwo  
\right\}^{-} ,
\end{align*}
where $\ms{D}^{-}$ denotes the closure of  $\ms{D}$. 
We set $ \ms{F}_{\pm \infty}^{0,0,0} :=
\{ z \Psi_{g} \, |\;  z \in \mbf{C} \} $. 
Let us define the  asymptotic in$/$out-going Fock space by  
  $ \ms{F}_{\pm \infty}=
\bigoplus_{n,l,m} \ms{F}_{\pm \infty}^{n,l,m} $. 
Let
\begin{align*}
\ms{F}^{n,l,m} =&
\ms{L}
\left\{
a_{r_{1} }^{\ast} (\xi_{1})
\cdots a_{r_{n} }^{\ast} (\xi_{n})
b_{s_{1}}^{\ast} (\eta_{1})
\cdots b_{s_{l}}^{\ast} (\eta_{l})
d_{\tau_{1}}^{\ast} (\zeta_{1})
\cdots d_{\tau_{m}}^{\ast} (\zeta_{m})
\Omega_{\el} \tens \Omega_{\ph} , \;  \right. \\  
 &| \left. \;  
 \xi_{i} \in \domain{\omega^{-1/2}}, \; 
 \eta_{j} , \; \zeta_{k} \in \eltwo  
\right\}^{-} .
\end{align*}
We define the wave operator by
$W_{\pm \infty}^{n,l,m}  : \ms{F}^{n,l,m} \to
\ms{F}_{\pm \infty}^{n,l,m}  $   by
\begin{align*}
W_{\pm \infty}^{n,l,m}  &
a_{r_{1}}^{\ast} (\xi_{1})
\cdots a_{r_{n} }^{\ast} (\xi_{n})
b_{s_{1}}^{\ast} (\eta_{1})
\cdots b_{s_{l}}^{\ast} (\eta_{l})
d_{\tau_{1}}^{\ast} (\zeta_{1})
\cdots b_{\tau_{m}}^{\ast} (\zeta_{m})
\Omega_{\el} \tens  \Omega_{\ph}  \\
&  \; :=
a_{r_{1}, \pm \infty}^{\ast} (\xi_{1})
\cdots a_{r_{n}, \pm \infty}^{\ast} (\xi_{n})
b_{s_{1}, \pm \infty}^{\ast} (\eta_{1})
\cdots b_{s_{l}, \pm \infty}^{\ast} (\eta_{l})
d_{\tau_{1}, \pm \infty}^{\ast} (\zeta_{1})
\cdots b_{\tau_{m}, \pm \infty}^{\ast} (\zeta_{m})
\Psi_{g} .
\end{align*}
By the commutation relations given by Lemma \ref{asympcommutation} and lemma \ref{asympanticommu}
$ W_{\pm \infty}^{n,l,m} $ can be extended
 to the unitary operator from $\;  \ms{F}^{n,l,m} \; $ onto
  $ \; \ms{F}_{\pm \infty}^{n,l,m} $.
Let $ \; W_{\pm \infty} = \oplus_{n,l,m} 
\overline{ W_{\pm \infty}^{n,l,m}  }   $.

$\quad $ \\ 
\textbf{(Proof of Theorem \ref{spectrumgap})}  \\
Let $\xi_{i} \in \domain{\omega^{-1/2}} , \; i=1, \cdots ,n $,
 $ \; \eta_{j} \in \eltwo , \; j=1, \cdots ,l $, and 
  $ \; \zeta_{k} \in \eltwo , k=1, \cdots ,m $.
 By Lemma \ref{Weylrelation},
 \begin{align*}
 &e^{itH} 
a_{r_{1}, \pm \infty}^{\ast} (\xi_{1})
\cdots a_{r_{n}, \pm \infty}^{\ast} (\xi_{n})
b_{s_{1}, \pm \infty}^{\ast} (\eta_{1})
\cdots b_{s_{l}, \pm \infty}^{\ast} (\eta_{l})
d_{\tau_{1}, \pm \infty}^{\ast} (\zeta_{1})
\cdots d_{\tau_{m}, \pm \infty}^{\ast} (\zeta_{m})
\Psi_{g}  \\
 &=e^{itE_{0} (H)}a_{r_{1} , \pm \infty}^{\ast} (e^{it \omega } \xi_{1}) \cdots a_{r_{n}, \pm \infty }^{\ast} (e^{it \omega } \xi_{n}) \\
& \quad \times 
b_{s_{1}, \pm \infty}^{\ast} (e^{it E_{M} } \eta_{1})
\cdots b_{s_{l}, \pm \infty}^{\ast} (e^{it E_{M}} \eta_{l})
d_{\tau_{1}, \pm \infty}^{\ast} (e^{itE_{M}} \zeta_{1})
\cdots d_{\tau_{m}, \pm \infty}^{\ast} (e^{it E_{M}} \zeta_{m})
\Psi_{g} . 
 \end{align*}
Then, $e^{itH}$ leaves $\ms{F}_{\pm \infty} $ invariant,
 and hence $H$ is reduced by $\ms{F}_{\pm \infty}  $. 
Then, 
 \begin{align*}
&W_{\pm \infty} e^{it H_{0}} 
a_{r_{1}}^{\ast} (\xi_{1}) 
\cdots a_{r_{n} }^{\ast} (\xi_{n} )
b_{s_{1}}^{\ast} (\eta_{1})
\cdots b_{s_{l}}^{\ast} (\eta_{l})
d_{\tau_{1}}^{\ast} (\zeta_{1})
\cdots d_{\tau_{m}}^{\ast} (\zeta_{m})
\Omega_{el} \tens  \Omega_{ph}  \\
 &= e^{it(H -E_{0}(H))} W_{\pm\infty} \;
a_{r_{1}}^{\ast} (\xi_{1}) 
\cdots a_{r_{n} }^{\ast} (\xi_{n} )
b_{s_{1}}^{\ast} (\eta_{1})
\cdots b_{s_{l}}^{\ast} (\eta_{l})
d_{\tau_{1}}^{\ast} (\zeta_{1})
\cdots d_{\tau_{m}}^{\ast} (\zeta_{m})
\Omega_{\el} \tens  \Omega_{\ph} .
\end{align*}
Thus, we obtain $ \; W_{\pm \infty} e^{it(H_{0} +E_{0}(H))} = e^{itH} W_{\pm \infty} $, 
on $\ms{F_{\pm \infty}}$. Then we have 
$ \; H_{0} + E_{0} (H) =  W_{\pm \infty}^{\ast}
H_{\restr \ms{F}_{\pm \infty} } W_{\pm \infty}  $.
 Thus, we obtain 
$ \sigma ( H_{0} + E_{0} (H)) \subset \sigma (H) $, and hence
$  [ E_{0}(H) , \infty ) \subset \sigma (H)$. 
On the other hand, it is trivial to see
$ \sigma (H) \subset [ E_{0}(H) , \infty )$. Hence, the proof is completed. $\blacksquare$

\section{Total Charge of Ground States}
It is seen that for $\eta, \zeta \in \eltwo$, 
\begin{align}
&[ N_{+} , b_{s}(\eta ) ] = - b_{s}(\eta ) , \qquad 
[ N_{+} , b^{\ast}_{s}(\eta ) ] = - b^{\ast}_{s}(\eta ) , 
\label{11/28.5} \\
&[ N_{-} , d_{\tau}(\zeta ) ] = - d_{\tau}(\zeta ) , \qquad 
[ N_{-} , d^{\ast}_{\tau}(\zeta ) ] = - d^{\ast}_{\tau} (\zeta ) ,
\label{11/28.6}
 \end{align}
on $ \ms{F}_{el}^{\fin} ( L^{2} (\mbf{R}^{3}  ; \mbf{C}^{4}) ) $.

\begin{lemma}   \label{11/28.4}
It follows that on $ \ms{F}_{el}^{\fin} ( L^{2} (\mbf{R}^{3}  ; \mbf{C}^{4}) ) $
\begin{equation}
  [Q,  \psi_{l} (\mbf{x} )^{\ast}  \psi_{l'} (\mbf{x} )  ] = 0 , 
  \qquad 
  [Q,  \psi_{l} (\mbf{x} )^{\ast} \alpha^{j} \psi_{l'} (\mbf{x} )  ] =0, 
\end{equation}
 for each $ \mbf{x}  \in \Rthree $.
\end{lemma}
\textbf{(Proof)} \\
By (\ref{11/28.5}) and (\ref{11/28.6}), it is seen that 
\[
 [N_{+}  , b^{\ast}_{s} (g^{l}_{s, \mbf{x}}) 
 d^{\ast}_{s'} (g^{l'}_{s', \mbf{x}})   ] =  b^{\ast}_{s} (g^{l}_{s, \mbf{x}}) 
 d^{\ast}_{s'} (g^{l'}_{s', \mbf{x}})  , \quad 
  [N_{+}  , d_{s} (h^{l}_{s, \mbf{x}}) 
b_{s'} (g^{l'}_{s', \mbf{x}})   ] =  - d_{s} (h^{l}_{s, \mbf{x}}) 
b_{s'} (g^{l'}_{s', \mbf{x}}) , 
\]
and
\[
[N_{+}  , b^{\ast}_{s} (g^{l}_{s, \mbf{x}})  
 b_{s'} (g^{l'}_{s', \mbf{x}})   ] =    
[ N_{+},  d_{s} (h^{l}_{s, \mbf{x}}) 
d_{s'}^{\ast} (h^{l'}_{s', \mbf{x}}) ] =0 .
\]
We also see that
\[
  [N_{-}  , b^{\ast}_{s} (g^{l}_{s, \mbf{x}}) 
 d^{\ast}_{s'} (g^{l'}_{s', \mbf{x}})   ] = 
 b^{\ast}_{s} (g^{l}_{s, \mbf{x}}) 
 d^{\ast}_{s'} (g^{l'}_{s', \mbf{x}}) , \quad 
[N_{-}  , d_{s} (h^{l}_{s, \mbf{x}}) 
b_{s'} (g^{l'}_{s', \mbf{x}})   ] = -d_{s} (h^{l}_{s, \mbf{x}}) 
b_{s'} (g^{l'}_{s', \mbf{x}}) , 
\]
and
\[
[N_{-}  , b^{\ast}_{s} (g^{l}_{s, \mbf{x}})  
 b_{s'} (g^{l'}_{s', \mbf{x}})   ]  = 
[ N_{-},  d_{s} (h^{l}_{s, \mbf{x}}) 
d_{s'}^{\ast} (h^{l'}_{s', \mbf{x}}) ] =0 .
\]
Hence, $  [N_{+},  \psi_{l} (\mbf{x} )^{\ast}  \psi_{l'} (\mbf{x} )  ] =
[N_{-},  \psi_{l} (\mbf{x} )^{\ast}  \psi_{l'} (\mbf{x} )  ]
= \sum_{s, s'}  (    b^{\ast}_{s} (g^{l}_{s, \mbf{x}}) 
 d^{\ast}_{s'} (g^{l'}_{s', \mbf{x}})     - d_{s} (h^{l}_{s, \mbf{x}}) 
b_{s'} (g^{l'}_{s', \mbf{x}})   ) $ follows. 
$\blacksquare$

$\quad$ \\
\begin{lemma} \label{12/1.1}
Assume \textbf{(A.1)}-\textbf{(A.3)}. Then  $e^{-itQ} $
 leaves $ \ms{D} (H) $ onto itself and 
\begin{equation}
e^{itQ \tens I } H  e^{-itQ \tens I} = H  ,
\end{equation}
 on $\ms{D} (H )$.
\end{lemma}
\textbf{(Proof)} \\
Let $\Psi \in \ms{D}_{0}$ with 
$
\Psi = a_{r_{1}}^{\ast} (\xi_{1})
\cdots a_{r_{n} }^{\ast} (\xi_{n})
b_{s_{1}}^{\ast} (\eta_{1})
\cdots b_{s_{l}}^{\ast} (\eta_{l})
d_{\tau_{1}}^{\ast} (\zeta_{1})
\cdots b_{\tau_{m}}^{\ast} (\zeta_{m})
\Omega_{\el} \tens  \Omega_{\ph} $.
It is trivial to see $ [Q \tens I , \; H_{0} ] = 0 $. 
For $ \Phi \in \ms{D} (H ) $, 
it is seen from Lemma \ref{11/28.4} that
\begin{align*}
&( \Phi ,   [ Q \tens I , \; H_{\I}' ] \;  \Psi )
= \int_{\Rthree}
\chi_{\I} (\mbf{x} ) 
 ( \Phi, \; 
[ \; Q , \; \psi^{\ast}(\mbf{x} ) \alpha^{j} \psi (\mbf{x} ) ]
  \tens  A_{j}(\mbf{x} ) \, \Psi )  \dx = 0,  \\ 
&  ( \Phi ,   [ Q \tens I , \, H_{\II}' ] \;  \Psi ) = 
=\int_{\Rthree \times \Rthree} 
\frac{\chi_{\II}(\mbf{x} )  \chi_{\II}(\mbf{y})}{|\mbf{x} - \mbf{y}|}
 ( \Phi , \; [ \; Q, \, 
\psi^{\ast}(\mbf{x} ) \psi (\mbf{x} ) \psi^{\ast}(\mbf{y})\psi (\mbf{y})   ]   \otimes I   \Psi )
  \dx \, \dy   = 0 .
\end{align*}
Hence, we obtain $ [Q \tens I  , \;  H  ] \Psi = 0 $. 
 Since $\Psi $ is an analytic vector of $Q$, we get 
\[
e^{itQ \tens I } H  e^{-itQ \tens I}  \Psi =
\sum_{n=1}^{\infty} \frac{(it)^{n}}{ n ! }
ad^{n}_{Q} H \Psi = H \Psi ,
\]
where $ \; ad^{0}_{Q}H := H $, $\; ad^{n}_{Q}H :=
 [Q , ad^{n-1}_{Q} H ], \; n\geq 1$. Since $\ms{D}_{0}$ is the core of $H$, we obtain
 $ e^{itQ \tens I } H  e^{-itQ \tens I}  \Psi = H \Psi $
 for $\ms{D}(H)$. $\blacksquare$

$\quad $  \\
\textbf{(Proof of Theorem \ref{TOTALCHARGEOFGS})} \\ 
From (\ref{ori1}) and (\ref{ori2}) it is seen that 
 for $ \Phi ,  \; \Psi  \in \domain{H}$, 
\begin{align}
& [ H'_{\I} , b_{s}(\eta ) ]^{0} (\Phi, \Psi )
= \int_{\Rthree}  \overline{\eta (\mbf{p} ) }
 (\Phi  , F_{+} (s, \mbf{p} )   \Psi ) 
d \mbf{p}  ,  \label{11/17.1}\\
 & [ H'_{\I} , d_{s}(\zeta ) ]^{0} (\Phi, \Psi )
= \int_{\Rthree}  \overline{\zeta (\mbf{p} ) }
 (\Phi  , F_{-} (s, \mbf{p} )   \Psi )  
d \mbf{p} ,  \label{11/17.2}
\end{align}
where
\begin{align}
& F_{+} (s, \mbf{p} ) \Psi 
= - \sum_{j,l,l'} \alpha_{l,l'}^{j}
\int_{\Rthree} \chi_{\I} (\mbf{x}) g_{s,\mbf{x}}^{l}  (\mbf{p})
 \;  \psi_{l'}(\mbf{x})   \otimes A_{j} (\mbf{x} ) \Psi \dx  ,
\label{Fplus} \\
&  F_{-} (s, \mbf{p} ) \Psi 
=  \sum_{j,l,l'} \alpha_{l,l'}^{j}
\int_{\Rthree} \chi_{\I} (\mbf{x}) h_{s,\mbf{x}}^{l'} (\mbf{p})
  \;  
\psi_{l}^{\ast} (\mbf{x})   \otimes A_{j} (\mbf{x} ) \Psi \dx 
\label{Fminus}  , 
\end{align}
and
\begin{align}
& [ H'_{\II} , b_{s}(\eta ) ]^{0} (\Phi, \Psi )
= \int_{\Rthree}  \overline{\eta (\mbf{p} ) }
 (\Phi  , G_{+} (s, \mbf{p} )   \Psi )_{\ms{F}_{\QED}}  
d \mbf{p}  ,  \label{11/17.3}\\
 & [ H'_{\II} , d_{s}(\zeta ) ]^{0} (\Phi, \Psi )
= \int_{\Rthree}  \overline{\zeta (\mbf{p} ) }
 (\Phi  , G_{-} (s, \mbf{p} )   \Psi )_{\ms{F}_{\QED}}  
d \mbf{p} ,  \label{11/17.4}
\end{align}
where 
\begin{align}
& G_{+} (s, \mbf{p} ) \Psi =
-\sum_{l} \int_{\Rthree \times \Rthree} 
\frac{\chi_{\II}(\mbf{x} )  \chi_{\II}(\mbf{y})}{|\mbf{x} - \mbf{y}|}
 \left( g_{s, \mbf{y}}^{l} (\mbf{p}) 
\rho (\mbf{x}) \psi_{l}(\mbf{y}) 
+  g_{s, \mbf{x}}^{l}  (\mbf{p})
 \; \psi_{l}(\mbf{x})  
 \rho (\mbf{y}) \right)  \tens I \; \dx \dy  \label{Gplus} \\ 
& G_{-} (s, \mbf{p} ) \Psi =
\sum_{l} \int_{\Rthree \times \Rthree} 
\frac{\chi_{\II}(\mbf{x} )  \chi_{\II}(\mbf{y})}{|\mbf{x} - \mbf{y}|}
 \left( h_{s, \mbf{y}}^{l} (\mbf{p}) 
\rho (\mbf{x}) \psi_{l}^{\ast}(\mbf{y}) 
+  h_{s, \mbf{x}}^{l}  (\mbf{p})
 \; \psi_{l}^{\ast}(\mbf{x})
 \rho (\mbf{y}) \right)  \tens I \; \dx \dy  . \label{Gminus}
\end{align}
Let $\Psi_{g}$ be the ground state of $H$. 
In a manner similar to \cite{Hi05}, we can obtain the 
 pull-through formula
 \begin{equation}
 | ( N_{\pm}^{1/2} \tens I ) \Psi_{g}  \|^{2}
 = \sum_{s=\pm 1/2}  \int_{\Rthree}  
\| ( H -E_{0}  + E_{M} (\mbf{p} )  )^{-1}
 \left(  \kappa_{\I} F_{\pm} (s, \mbf{p} ) + 
 \kappa_{\II} G_{\pm} (s, \mbf{p} ) \right)
\Psi_{g}   \|^{2} d \mbf{p}  . \label{pullplusminus}  
 \end{equation}
It is seen that from \textbf{(A.3)} and (\ref{AjHbound})
\begin{align}
 &\| F_{+} (s, \mbf{p} ) \Psi_{g} \| 
 \leq 
 \sum_{j,l,l'} | \alpha_{l,l'}^{j} | |g_{s,\mbf{0}}^{l} (\mbf{p}) |  
 \| \chi_{\I} \|_{L^{1}}  M_{l'} 
( L_{I}^{j}(\epsilon )  |E_{0}(H)| + 
R_{I}^{j}(\epsilon )    )\|  \Psi_{g} \|, \label{evaluateFplus}
\\
 &\| G_{+} (s, \mbf{p} ) \Psi_{g} \| 
 \leq  2 \sum_{\nu  , l}
  | g_{s, \mbf{0}} (\mbf{p}) |
M_{\II}   (M_{\nu}^{\el} )^{2} M_{l}^{\el} \|  \Psi_{g} \| 
 \label{evaluateGplus}
\end{align}
Then, there exist constants $\mu_{+} (s) >0$ and 
$ \nu_{+} (s) > 0 $ such that 
\begin{equation}
\| ( H -E_{0} (H) + E_{M} (\mbf{p} )  )^{-1}
 \left(  \kappa_{\I} F_{+} (s, \mbf{p} ) + 
 \kappa_{\II} G_{+} (s, \mbf{p} ) \right)
\Psi_{g}      \|
\leq 
(\kappa_{\I} \mu_{+} (s)    + \kappa_{\II} \nu_{+} (s))
  \frac{|u_{s}^{l} (\mbf{p})  |}{ \sqrt{E_{M} (\mbf{p})}^{3} } 
\| \Psi_{g}\| .
 \label{11/17.7}
 \end{equation}
Similarly, there exist constants $\mu_{-} (s)>0$ and 
$ \nu_{-} (s) > 0 $ such that 
\begin{equation}
\| ( H -E_{0} (H) + E_{M} (\mbf{p} )  )^{-1}
 \left(  \kappa_{\I} F_{-} (s, \mbf{p} ) + 
 \kappa_{\II} G_{-} (s, \mbf{p} ) \right)
\Psi_{g}      \|
\leq  
(\kappa_{\I} \mu_{-}  (s)  + \kappa_{\II} \nu_{-} (s) )
  \frac{|v_{s}^{l} (\mbf{p})  |}{ \sqrt{E_{M} (\mbf{p})}^{3} } 
\| \Psi_{g}\|.
 \label{11/17.8}
\end{equation}
Note that $  \domain{\sqzf{E_{M}}} \subset
\domain{N_{\pm} } $.  By (\ref{pullplusminus}), 
 (\ref{11/17.7}) and (\ref{11/17.8}), for sufficiently small 
 $\kappa_{\I}$ and  $\kappa_{\II}$,  it follows that
 \begin{equation}
 (\Psi_{g}  , N_{+} \Psi_{g})
  +  (\Psi_{g}  , N_{-} \Psi_{g}) < 1.
  \label{NplusNminus}
 \end{equation}
Now let us consider $\Psi_{g} \in \ms{F}_{n} $, $n \ne 0$.
 Then it follows that 
\begin{equation} 
(\Psi_{g}  , N_{+} \Psi_{g})
  -  (\Psi_{g}  , N_{-} \Psi_{g}) \geq 1  \qquad 
\text{o}r
 \qquad
  (\Psi_{g}  , N_{+} \Psi_{g})
  -  (\Psi_{g}  , N_{-} \Psi_{g})  \leq -1   .
    \label{contradictionNplusNminus}
\end{equation} 
 But this contradicts (\ref{NplusNminus}). 
Hence, $ \Psi_{g} \in \ms{F}_{0}$ follows.  $\blacksquare$

$\quad $\\
$\quad $\\
{\Large Appendix} \textbf{(Uniqueness of Ground States;
\cite{Hi05} )}  \\
Let $\ms{K}$ be a Hilbert space. We consider an abstract Hilbert space $\ms{H}$ as
\[
\ms{H} = \ms{K} \otimes \ms{F}_{\bos} ( L^{2}(\Rthree ; 
\mbf{C}^{2}) ).
\]
Let
\[
X_{0} = K \otimes I + I \otimes H_{\ph},
\]
and
\[
X (q ) = X_{0} + q X' , \quad \quad \quad q \in \mbf{R} , 
\]
The operator $K$ satisfies the following conditions:
\begin{quote}
\textbf{(H.1)} 
The operator $K$ is self-adjoint and bounded from below. 
\end{quote}

\begin{quote}
\textbf{(H.2)}  
 $X'$ is a symmetric operator on $\ms{H}$, and there exist constants  $ a>0  $ and $b>0$ such that 
\[
\| X' \Psi \| \leq a \| X_{0} \Psi \| + b \| \Psi \| , \quad \quad 
\Psi \in \ms{D} (H_{0}) .  
\]
\textbf{(H.3)} 
There exists an operator
 $ S(r,\mbf{k} ) : \ms{H} \to \ms{H} , \; \mbf{k} \in \mbf{R}^{3} , \;   r=1,2  $,  
 such that for $ \Phi , \; \Psi \in \ms{D} (H_{0}) $,
\[
( I \otimes a_{r}^{\ast} (f) \Phi , X' \Psi ) 
- ( X' \Phi , I \otimes a_{r} (f) \Psi ) 
=\int_{\Rthree} \overline{f(\mbf{k} )}( \Phi , S(r,\mbf{k} ) \Psi ) d\mbf{k} .
\] 
$\quad$ \\
\end{quote}
Assume that $\; X(q) =X_{0}+ q X' \;$ has a ground state
 $ \Psi_{0}(q)$ : $ X(q ) \Psi_{0}(q ) = E_{0} (X(q ))\Psi_{0}(q) $.
\begin{quote}
\textbf{(H.4)} 
Let $\Phi \in \ms{D} (X_{0})$. Then for $f \in C^{\infty}(\Rthree )$,  $ \; S(r, \mbf{k} ) $ in \textbf{(H.3)} satisfies 
\[
 \int_{\Rthree} \overline{f(\mbf{k} )} \left( \Phi , 
 e^{ it (X(q)-E_{0}(X (q )) + \omega (\mbf{k} )) } S(r,\mbf{k} ) \Psi_{0}(q) \right) d\mbf{k} 
\in L^{1}(  [ 0, \infty )  , \; dt ) ,
\]
and
$ \int_{\Rthree} \| S(r, \mbf{k} ) \Psi_{0}(q ) ) \|^{2} 
d \mbf{k} \,  < \, \infty  . $
\end{quote}
$\quad$ \\
\textbf{Theorem A.1 (\cite{Hi05};Theorem 2.9)}
Assume that \textbf{(H.1) - (H.4)}. Let
$\Psi_{0} (q) \;$ be an arbitrarily ground state.  
Then (a) and (b) are equivalent. \\
(a) $ \quad \Psi_{0} (q)  \in \ms{D} (I\otimes N_{b}^{1/2} )$. \\
(b) 
$ \quad \int_{\Rthree}  \| (X(q ) -E_{0} (X(q) ) + \omega (\mbf{k} ) )^{-1}
S(r, \mbf{k} ) \Psi_{0} (q ) \|^{2} d \mbf{k} < \infty  $. \\
In particular, if (a) or (b) holds, 
\[
\| ( I \otimes N_{\bos}^{1/2} ) \Psi (q) \|^{2} = 
q^{2} \sum_{r=1,2}  \int_{\Rthree}  \| (X(q ) -E_{0} (q ) + \omega (\mbf{k} ) )^{-1}
S(r, \mbf{k} ) \Psi_{0} (q ) \|^{2} d \mbf{k} .
\]

$\quad$ \\ 
\begin{quote}
\textbf{(H.5) (Spectral gap of K) }
$\quad   \inf \sigma_{ess} (K) - E_{0} (K)  > 0 $. 

$\quad$ \\
\textbf{(H.6)} 
Let $\ms{N}_{q} = \text{ker} ( X(q) -E_{0} (X(q)))$. Then
 it follows that
\[
\lim_{q \to 0}  \sup_{\Psi (q )  \in N_{q} \backslash
 \{ 0 \} }
 q^{2} \sum_{r=1,2}  \int_{\Rthree}  \| (X(q ) -E_{0} (q ) + \omega (\mbf{k} ) )^{-1}
S(r, \mbf{k} ) \Psi_{0} (q ) \|^{2} 
  d \mbf{k} / \| \Psi_{0} (q ) \|^{2}
  = 0   .
\]
\end{quote}

$\quad $ \\
\textbf{Theorem A.2 (\cite{Hi05};Theorem 4.2)} \\
Assume that  \textbf{(H.1) -(H.6)}.
Then there exists a constant $\tilde{q} >0 \; $ such that for 
$ |q| <  \tilde{q} $, 
\[
\text{dim  ker}(  X(q) -E_{0} (X_{q}) ) \leq 
\text{dim   ker} (K -E_{0}(K)) .
\]

$\quad $ \\
$\quad $ \\
{\Large Acknowledgments} \\
It is pleasure to thank Prof. F. Hiroshima for his advice and discussions.

\end{document}